\renewcommand{\theequation}{\thesection.\arabic{equation}}

\long\def\query#1{\hskip 0pt{\vadjust{\everypar={}\small\vtop to 0pt{\hbox{}%
     \vskip -13pt\rlap{\hbox to 49.0pc{\hfil{\vtop{\hsize=8pc\tolerance=6000%
     \hfuzz=.5pc\rightskip=0pt plus 3em\noindent#1}}}}\vss}}}}%

\def\start{\documentclass[12pt]{article}
                   \usepackage{amssymb,amsfonts,amsmath,amscd,latexsym,epsf}
                 \setlength{\textwidth}{17cm} \setlength{\textheight}{25.2cm}
 \hoffset -20mm \topmargin= -13mm \begin{document}%\version\versionno
}

\def \ignore#1 { {} } 

\def \bs {\bigskip}

\def \Fig#1#2#3 {
\begin{figure}
\begin{center}
\scalebox{.6}{\includegraphics{#1.eps}}
\label{#1}
\end{center}
\caption{#3}
\end{figure}
}

\def \en#1 {\ensuremath{#1}\ {}}

\def \H  {\en{H_3} }
\def \Hp {\en{H_3} }
\def \SL {\en{SL(2,\R)} }
\def \SU {\en{SU(2)} }
\def \AAA {\en{AdS_3} }
\def \AA {\en{AdS_2} }
\def \SLC {\en{SL(2,\C)} }
\def \SLU {\en{SL(2,\R)/U(1)} }
\def \SUU {\en{SU(2)/U(1)} }
\def \S   {\en{S^2} }
\def \sl {\en{sl(2,\R)} }
\def \su {\en{su(2)} }
\def \au {\en{\widehat{u}(1)} }
\def \asu {\en{\widehat{su}(2)} }
\def \asl {\en{\widehat{sl}(2,\R)} }
\def \aslc {\en{\widehat{sl}(2,\C)} }

\def \LL { L_0+\bar{L}_0-\frac{c}{24} }

\def \cyl {{\rm cyl}}
\def \cig {{\rm cigar } }
\def \tr  {{\rm trumpet }  }
\def \bell  {{\rm bell } }
\def \sign {{\rm sign }}
\def \cst {{\rm constant}}

\def \half {\frac{1}{2}}
\def \halfpi {\frac{\pi}{2}}
\def \halfi {\frac{i}{2}}
\def \halfib {\frac{i}{2b^2}}
\def \ip {i\frac{\pi}{2}}
\def \rr {\frac{r}{\pi b^2}}
\def \shift {\frac{i}{2b^2}}

\def \j {{-\half+iP}}

\def \ap {\alpha'}

\def \p {\partial}

\def \om {\omega}
\def \g {\gamma}
\def \bg {\bar{\gamma}}
\def \vf {\varphi}

\def \rar {\rightarrow}

\def \bz {\bar{z}}
\def \bp {\bar{\partial}}

\def \Tr {{\rm Tr }\ }

\def \Z {\mathbb{Z}}
\def \N {\mathbb{N}}
\def \R {\mathbb{R}}
\def \C {\mathbb{C}}

\def \la {\left\langle}
\def \ra {\right\rangle}
\def \lla {\langle\langle}
\def \rra {\rangle\rangle}

\def \o {_{os}}

\def \tq {\tilde{q}}
\def \tth {\tilde{\theta}}
\def \ttau {\tilde{\tau}}

\def \pp {_{k\rightarrow \infty}}

\def \HL {{\cal H}_L{}}
\def \HR {{\cal H}_R{}}
\def \HT {{\cal H}}

\def \vt {\vartheta_1}

\def \pieta {\Pi_{n=1}^\infty (1-q^n)}

\def \bea {\begin{eqnarray}}
\def \eea {\end{eqnarray}}
\def \ber {\begin{eqnarray*}}
\def \eer {\end{eqnarray*}}
\def \nn  {\nonumber}
\def \be  {\begin{equation}}
\def \ee  {\end{equation}}

\def\N{{\cal N}}
\def\SN{N}
\def\Dz{{\rm D0}}
\def\Do{{\rm D1}}
\def\D2{{\rm D2}}
\def\s{\sigma}
\def\U{{\rm U(1)}}
\def\a{\alpha}
\def\cJ{{\cal J}}

\def \Ht {\en{H_2} } 
\def\tz{\tilde z}
\title{\bf Branes in the 2D black hole}  
\author{\\[5mm] Sylvain Ribault$^1$ and Volker Schomerus$^2$ 
  \\[5mm]$^1$  Department of Mathematics, King's College London \\
Strand, London WC2R2LS,
United Kingdom
\\[3mm] 
$^2$ Service de Physique Th{\'e}orique, CEA Saclay,\\ 
F-91191 Gif-sur-Yvette CEDEX, France\\[5mm] }
\documentclass[12pt]{article}
                   \usepackage{amssymb,amsfonts,amsmath,amscd,latexsym,epsf,graphicx}
\begin{document}%\version\versionno
%\start
\renewcommand{\theequation}{\thesection.\arabic{equation}}

\begin{titlepage}      \maketitle       \thispagestyle{empty}

\vskip1cm
\begin{abstract} 
We present a comprehensive analysis of branes in the Euclidean 
2D black hole (cigar). In particular, exact boundary states and 
annulus amplitudes are provided for D0-branes which are localized 
at the tip of the cigar as well as for two families of extended D1 
and D2-branes. Our results are based on closely related studies for 
the Euclidean \AAA model \cite{pst} and, as predicted by the  
conjectured duality between the 2D black hole and the sine-Liouville
model, they share many features with branes in Liouville theory. 
New features arise here due to the presence of closed string modes 
which are localized near the tip of the cigar. The paper concludes 
with some remarks on possible applications.      
\end{abstract} 

\vspace*{-17.9cm}\noindent {\tt SPhT-T03/146} 
\\ 
  {\tt CPHT-RR 045.0803}\\
  { \tt {KCL-MTH-03-15}}
\bigskip\vfill
\noindent
\phantom{wwwx}{\small e-mail: }{\small\tt
ribault@mth.kcl.ac.uk, vschomer@spht.saclay.cea.fr} 

\end{titlepage} 
\baselineskip=19pt

\tableofcontents
\def\SLR{\SL}

\setcounter{equation}{0} 
\section{Introduction} 

There exists an interesting exact 2D classical solution of the 
string equations of motion which became known as the {\em cigar}
and which has been studied for various different reasons after it 
was first discovered in \cite{EFR,MSW}. Initially, investigations 
focused mainly on a Lorentzian version of the cigar geometry 
because is was observed to exhibit all characteristic features of 
a 1+1 dimensional black hole \cite{Wit}. The Euclidean counterpart, 
which we shall mostly refer to as {\em 2D black hole}, gained more 
relevance later as a building block for several important 9+1-dimensional 
string backgrounds \cite{HoSt,OoVa}. In particular, it arises when a 
stack of NS5-branes is separated along a 1-dimensional circle 
\cite{sfetsos} in order to prevent the string coupling constant 
from diverging at the branes' locus \cite{OoVa}. Hence, a detailed 
study  of strings and branes in the 2D black hole geometry can be 
seen as a crucial step toward understanding string theory in the 
near horizon geometry  of NS5-branes and its conjectured duality 
with little string theory (see e.g. \cite{Kutasov:2001uf} and 
references therein). 
\smallskip

The spectrum of closed string modes in the 2D black hole geometry 
was originally analyzed by Dijkgraaf and the Verlindes \cite{dvv}. 
Even though their proposal was partly based on a conjectural and 
somewhat obscure duality with the so-called 2D trumpet geometry, 
it has been confirmed recently through an exact computation
of the partition function \cite{hpt}.\footnote{There is a only a 
small discrepancy concerning the so-called 
unitarity bound.} What makes the whole structure of the spectrum 
non-trivial is the existence of closed string modes which are 
localized near the tip of the cigar. Other exact quantities in the 
bulk theory, in particular the closed string 3-point couplings, can 
be inferred from Teschner's work on the \H model \cite{teschnerh3,
joergfusion}. The latter is related to the \SL WZW model by a Wick 
rotation in target space and it serves as the starting point for a 
coset construction of the 2D black hole.
\smallskip 

In this work we extend the exact solution of string 
theory in the 2D black hole background to branes and open strings. 
One obvious motivation comes from the duality with little string 
theory that we have mentioned before. It is also worth pointing out 
that the 2D black hole was conjectured by Fateev, Zamolodchikov 
and Zamolodchikov to be  T-dual to sine-Liouville theory (see 
\cite{KaKoKu} for a more precise description of the conjecture). 
A supersymmetric version of this T-duality \cite{doubscal} which 
involves the $N=2\ $ \SLU Kazama-Suzuki quotient on one side 
and $N=2$ Liouville theory on the other has been firmly established
\cite{HorKap}. In this light, our studies of branes and open strings 
in the 2D black hole appear as a 2-dimensional generalization of 
similar investigations for Liouville theory \cite{zzf,zzb}. We 
shall find many parallels between branes in the two models.  
\smallskip 

Our construction of branes and their open string spectra in the 
2D black hole background departs from the results of
\cite{pst,deuxbranes} on D-branes in \H 
(see also \cite{schwimmer,RajRoz,plo} for some proposals and  
partial results on exact boundary theories for branes in \H). 
In principle, these descend down from the 3-dimensional \H model 
through a rather simple coset procedure involving some 1-parameter 
family of translations. The implications for the 2-dimensional 
quotient, however, are quite non-trivial due to the discrete bulk 
modes that are localized near the tip of the 2D black hole. Note 
that these have no analogue in the bulk spectrum of the \H-model. 
Hence, e.g.\ the Cardy-type consistency conditions which arise 
from world-sheet duality differ significantly 
between the parent theory and its coset. 
\smallskip 
  
We shall start our presentation with a review of some relevant 
material on the bulk theory. This is followed by three rather 
independent sections on three different types of branes. 
The latter may be classified according to 
their dimension and they include a set of point-like branes 
as well as two families of extended D1 and D2-branes. Each 
section below contains a careful study of the semi-classical 
limit in which we describe the branes' geometry and some 
properties of their open string spectra\footnote{A Born-Infeld
analysis of branes in the 2D black hole has been worked out 
previously in \cite{angelos}.}. A comprehensive summary of the 
exact solutions is provided and motivated from the results of 
\cite{pst} before we explain how to verify the consistency with 
world-sheet duality.\footnote{The solution for the D1-branes was  
already presented in \cite{pst} and \cite{RibPhD}.} The paper 
concludes with a list of open problems.

\setcounter{equation}{0} 
\section{Preliminaries: Closed strings on the cigar}

To prepare for our study of branes and open strings in the cigar 
geometry we will need to review some background material about the closed 
string theory. We will start with a few words on the minisuperspace model, 
partly to motivate the description of the spectrum in the full conformal 
field theory which we address in the second subsection. The section 
concludes with a short review of the coset construction from \H.

\subsection{The minisuperspace model} 

\Fig{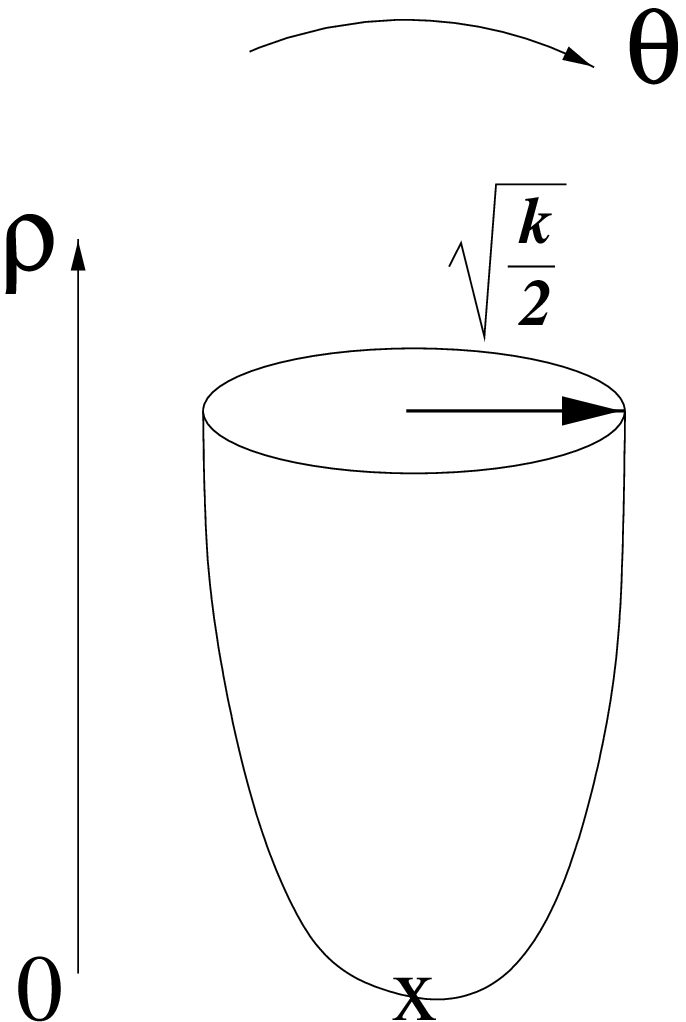}{5}{The 2d black hole}

The 2-dimensional cigar is parametrized by some angle $\theta \in [0,2\pi[$ 
and a radial coordinate $\rho \in [0,\infty[$ with $\rho = 0$ 
corresponding to the tip (see figure \ref{71cig}). 
In these coordinates, the background 
metric and dilaton are given by the expressions 
\bea
ds^2 \ =\   \frac{k}{2} \, (d\rho ^2+\tanh^2\rho\,  d\theta ^2) 
\ \ \ \ \ , \ \ \ \ \ \  
e^\Phi \ = \   \frac{e^{\Phi_0}}{\cosh \rho}   
\eea
where we have used the string frame and set $\ap=\half$. We can 
get some intuition into the spectrum of closed string modes from 
the minisuperspace approximation. To this end, we are looking for
eigen-functions of the Laplacian on the cigar, 
\bea
\Delta  & = &  
- \frac{1}{e ^{-2\Phi}\sqrt{\det
  G}}\ \partial _\mu e ^{-2\Phi}\sqrt{\det G}\ 
  G^{\mu \nu}\partial_\nu \nonumber \\[2mm] & = & 
-  \frac{2}{k}\left[  \partial_\rho^2 + (\coth \rho +
  \tanh \rho)\, \p_\rho+\coth^2\rho\, \p_\theta ^2 \right]\ \ . 
\eea  
The $\delta$-function normalizable eigen-functions of this operator 
can be expressed in terms of hypergeometric functions through 
\bea
\phi^j_{n0}(\rho,\theta) & = & -\frac{\Gamma(-j+\frac{|n|}{2})^2}
  {\Gamma(|n|+1)\Gamma(-2j-1)} \ e^{in\theta}\sinh ^{|n|}\rho\
  \times \nonumber \\[2mm] & & \hspace*{2cm} \times \ 
  F\left(j+1+\frac{|n|}{2},-j+\frac{|n|}{2}, |n|+1;-\sinh^2\rho\right)
\label{wavecig}
\eea 
where $j \in -1/2 + i \mathbb{R}$ describes the momentum along 
the $\rho$-direction of the cigar and $n \in \mathbb{Z}$ is the 
angular momentum under rotations around the tip. For the 
associated eigenvalues one finds 
\bea \label{EV}
\Delta ^j_{n0}\ =\ -\frac{2j(j+1)}{k}+\frac{n^2}{2k}\ \ \ .
\eea   
In the symbols $\phi^j_{n0}$ and $\Delta ^j_{n0}$  we have inserted 
an index `0' without any further comment. Our motivation will become 
clear once we start to look into the spectrum of the bulk CFT. Let us
also stress that there are no $L^2$-normalizable eigenfunction of the 
Laplacian on the cigar. Such eigenfunctions would correspond to {\em 
discrete states} living near the tip, but in the minisuperspace 
approximation one only finds exclusively {\em continuous states} 
which behave like plane waves at $\rho\rightarrow \infty$.

\subsection{The spectrum of the bulk CFT} 
We now turn to the discussion of the conformal field theory (CFT) on 
the cigar which is known to possess a central charge $c = 2(k+1)/(k-2)$. 
Each of the wave functions (\ref{wavecig}) in the minisuperspace theory 
lifts to a {\em primary field} in this conformal field theory. But the 
full story must be a bit more complicated. In fact, at $\rho \rightarrow 
\infty$, the cigar looks like an infinite cylinder and for the latter 
we know that primary fields are labeled by momentum $n$ and winding 
$w$ around the compact circle, in addition to the continuous 
momentum $iP = j+\half$ along the uncompactified direction. Hence, 
we expect that the full conformal field theory on the cigar 
has primary fields $\Phi^j_{nw}(z,\bz)$. Obviously, the winding 
modes $w \neq 0$ do not show up in the particle limit so that the
minisuperspace analysis provides no help in deciding which values 
$n,w$ and $j$ can run through. The answer, however, is known and 
goes back to the work of Dijkgraaf, Verlinde and Verlinde 
\cite{dvv}. It turns out that the closed string spectrum is made of
two different series one of which is the {\em continuous series} 
with 
$$ j \ \in\  -\frac12 + i\, \mathbb{R}^+_0 \ \ \ \ \ , \ \ \ \ \ 
   n \ \in \ \mathbb{Z} \ \ \ \ , \ \ \ \ w \ \in \ \mathbb{Z} \ \ . 
$$ 
This is the series we have argued for above. But in addition 
there is also a {\em discrete series} of primary fields for 
which $w,n \in \Z$ are such that 
\bea
j \, \in \, \cJ^d_{nw} \ :=\  \left[\, \frac{1-k}{2}\, ,\, 
- \frac12 \, \right] \, \cap  
\, \left( \mathbb{N} - \half |kw| + \half |n| \right)  \ . \label{rangej}
\eea      
This set of primary fields was first 
described in \cite{dvv} with a slightly different lower 
bound on $j$. The bound we have written here appears in 
\cite{hpt} (see also \cite{moi} for a related bound in 
the case of strings on \SL). We shall confirm it later 
through world-sheet duality involving open strings. Let us
also note that for $w = 0$, the discrete series is empty, 
in agreement with the minisuperspace analysis. The primary 
fields of these two series have conformal weights given by        
\bea \label{h}
h^j_{n w} \ = \  - \frac{j(j+1)}{k-2} + \frac{(n+kw)^2}{4k} 
   \ \ \ \mbox{ and } \ \ \ 
   {\bar h}^j_{n w} \ = \ 
    -\frac{j(j+1)}{k-2}+\frac{(n-kw)^2}{4k} 
\ \ . 
\eea
Notice that these conformal weights are all positive, as for 
any Euclidean unitary conformal field theory.
In the limit of large level $k$, the sum $h^j_{n0} + \bar h^j_{n0}$ 
of the left and right conformal weights with $w=0$ reproduces the 
spectrum (\ref{EV}) of the minisuperspace Laplacian. 
\smallskip

Over each of these primary fields there is a whole tower of 
descendants. The spectrum of conformal weights appearing in these
towers is encoded in products of left- and right-moving chiral 
characters. For the continuous series, the chiral characters 
read as follows
\bea \label{carcont}
\chi^c_{(j,\om)}(q)\ =\ {\rm Tr}_{(j,\om)} \ q^{L_0-\frac{c}{24}}
      \ = \ \frac{q^{-\frac{(j+\frac{1}{2})^2}{k-2}+
     \frac{\om^2}{k}}}{\eta(q)^2}\ \ .
\eea
Here $\omega$ is some real number parametrizing the \SLU\ chiral
algebra representations descending from a continuous representation of
\SL\ of spin $j=-\half +iP$. In the case of the discrete series, the 
expressions for chiral characters are a bit more complicated (see 
also Appendix A), 
\bea\label{cardisc}
\chi_{(j,\ell-j)}^{d}(q) & = & \frac{q^{-\frac{(j+\frac12)^2}{k-2}+
 \frac{(j-\ell)^2}{k}}}{ \eta(q)^2}\ 
 \left[\, \epsilon_\ell\, 
\sum_{s=0}^\infty \ (-1)^s q^{\half
  s(s+2|\ell+\half|)} -\frac{\epsilon_\ell-1}{2}\,
  \right]\ \ \\[3mm]
& & \hspace*{.5cm}\mbox{ where } \ \ \ \epsilon_\ell \ =\ 
\left\{ \begin{array}{rcl}
1 & {\rm if } & \ell \geq \ 0 \\[1mm]
-1  & {\rm if } & \ell \leq -1
\end{array}\right. \ \ . \nonumber 
\eea
Our formula for $\chi^d$ agrees with results in \cite{Pakman} as
long as $\ell \geq 0$ (modulo the redefinition $j\rar -j$).% 
\smallskip

Out of the characters we have introduced, one can build the full 
partition function of the bulk theory as follows 
\bea 
   Z(q,\bar{q})  &=&  \sum_{n,w\in \Z}\ \int dj\ N(j,n,w) 
   \ \chi^c_{(j,(n+kw)/2)}(q) \  
                           \chi^c_{(j,(-n+kw)/2)}(\bar{q})   
\\[2mm]   
 &+& \  
\sum_{n,w\in \Z}\, \sum_{j \in \cJ^d_{nw}} 
  \  \chi^d_{(j,(n+kw)/2)}(q)\, \chi^d_{(j,(-n+kw)/2)}(\bar{q}) \ \ . 
\eea  
Here, $N(j,n,w)$ is some nontrivial density of states for the
continuous representations which can be found in \cite{hpt}. 
Comparison of the conformal weights (\ref{h}) with the chiral characters 
(\ref{carcont}) and (\ref{cardisc}) shows that both in the 
discrete and in the continuous sectors, the closed string 
quantum numbers $n,w$ are related to the left and right chiral 
representation labels $\om,\bar \om$ through
\bea
\om\ =\ \frac{n+kw}{2}\ \ \ , \ \ \ \bar \om\ =\ \frac{-n+kw}{2}\ \ .
\eea
Before we conclude this subsection let us remark that there exists 
also explicit formulas for the operator product expansions of the 
primary fields. We will not need them here in full generality and 
refer the interested reader to the literature
%SR added refs
(for instance, \cite{becker,teschnerh3,gehenne,moiii}). 
We will need, however,  
the following explicit expression for the two-point functions of 
primary fields 
\bea 
\la \Phi ^j_{nw}(z,\bar z)\, \Phi^{j'}_{n'w'}(w,\bar w) \ra \ =\ 
  \left[\delta(j+j'+1)+ {\cal R}(j,n,w)\, 
 \delta(j-j') \right] \ \frac{\delta_{n+n'}\delta_{w+w'}}
  {|z-w|^{2 h^j_{nw} + 2h^j_{nw}}} 
\label{twopt}
\eea
with 
\bea \label{R}
{\cal R}(j,n,w)  \ =\   \nu_b^{2j+1}\ \frac{
\Gamma(2j+1)}{\Gamma(-2j-1)}\ 
  \frac{\Gamma(-j+\frac{n-kw}{2})\Gamma(-j+\frac{n+kw}{2})} 
  {\Gamma(j+1+\frac{n-kw}{2})\Gamma(j+1+\frac{n+kw}{2})}
 \ \frac{\Gamma(1+b^2(2j+1))}{\Gamma(1-b^2(2j+1))}\ \ 
\eea
and the standard notations 
\bea
b^2\ =\ \frac{1}{k-2}\ \ \ , \ \ \
\nu_b \ =\ \frac{\Gamma(1-b^2)}{\Gamma(1+b^2)}\ \ .
\eea
The quantity (\ref{R}) is known as {\em reflection amplitude} since it 
describes the behavior of the fields $\Phi^j_{nw}$ under the reflection 
$j \rightarrow -j-1$, 
$$ \Phi^j_{nw}(z,\bar z) \ = \ {\cal R}(j,n,w) \ 
    \Phi^{-j-1}_{nw}(z,\bar z)\ \ . 
$$  
Observe that the bulk two-point functions become singular for fields 
from the discrete series because of the factors $\Gamma (-j+
\frac{n\pm kw}{2})$. This singularity is caused by the specific 
normalization we have chosen in which the discrete fields are 
obtained from the continuous series through analytic continuation. 
Working with such a normalization is not really problematic since the
divergent terms will cancel each other in all physical quantities. Let 
us also point out that due to the presence of the $\delta$-function, 
the two-point functions are infinite whenever $j = j'$. The origin of
this divergence is obvious: it is related to the infinite volume of 
the cigar and can be regularized by introducing a cut-off $V_0$. We
will not do that explicitly, but eventually it is important to keep 
the issue in mind.

\subsection{Coset construction from \H } 

Our aim here is to outline how the quantities of the previous 
subsection can be recovered from the theory on the 3-dimensional 
Euclidean space \H $\sim$ \SLC/\SU. 
The latter is related to \SL\  by a Wick rotation. An exact 
solution of this closed string background has been proposed in 
\cite{teschnermini,teschnerh3} and it has subsequently been 
verified in \cite{Teschnercross}. Since the cigar geometry 
can be constructed as a coset $\H/\mathbb{R}$ of $\H$ the two 
conformal field theories are certainly closely related, though  
there exist some important subtleties. 
\smallskip     

\Fig{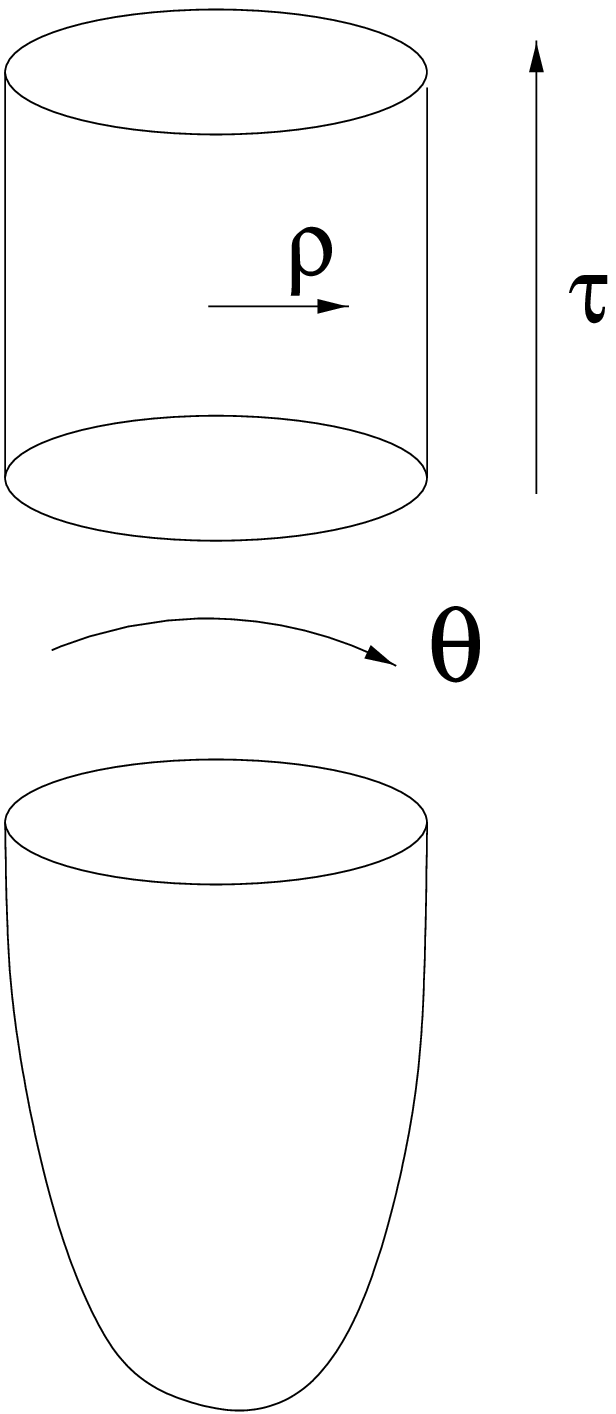}{3}{The cigar from \H }

Let us first understand how the descent from $\H$ to the cigar 
works in the framework of the minisuperspace model. On $\H$ we 
use a third coordinate $\tau \in \mathbb{R}$ in addition to the 
two coordinates $\rho$ and $\theta$ we have used for the cigar
(see figure \ref{72cigads}). The metric on $\H$ takes the form   
$$
 ds^2\ =\ \frac{k}{2}\, (d\rho ^2 + \cosh ^2\rho\, d\tau ^2 + 
      \sinh ^2\rho \, d\theta ^2) \ \ . 
$$ 
Furthermore, there is a non-trivial B-field and the dilaton on 
$\H$ is constant. Using this information we can write down the 
Laplacian $\Delta^{\rm H}$ of the minisuperspace theory, 
$$ 
\Delta^{\rm H} \ = \ - \frac{2}{k}\, \left[\partial_\rho^2 + 
 (\coth \rho + \tanh \rho)\,  \partial_\rho + {\cosh^{-2}\rho}\, 
\partial_\tau^2 + {\sinh^{-2}\rho}\,  \partial_\theta^2\right] 
\ \ . 
$$ 
Eigen-functions $\phi^{{\rm H},j}_{n,p}$ of this operator are 
labeled by three quantum numbers $(j,n,p)$ where $p\in \mathbb{R}$ 
denotes the momentum along the $\tau$-direction, i.e.\ the 
eigen-value of the operator $i\p_\tau$. Explicit formulas for 
the eigen-functions of $\phi^{\rm H}$ are easy to find (see e.g.\ 
\cite{pst}). From the following formula for the difference between 
the Laplacians on $\H$ and the cigar  
$$ \Delta^{\rm H} - \Delta \ = \ \frac{2}{k}\, \left[-\frac{1}{\cosh^2\rho} 
\, \partial_\tau^2  + \partial^2_\theta\right] \ \ 
$$
we can read off that eigen-functions and eigen-values of the Laplacians
in the two backgrounds are related by 
\bea \label{eigencigar}
\phi^j_n (\rho,\theta) \ = \ \phi^{{\rm H}, j}_{n,p=0} (\rho,\theta,\tau)
\ \ \ \ \ , \ \ \ \ \ 
\Delta^j_{n,0} \ = \ \Delta^{{\rm H}, j}_{n,p=0} + \frac{n^2}{2k}\, \ = \ 
-\frac{2j(j+1)}{k} + \frac{n^2}{2k} \ \ .                       
\eea 
Note that the formula for the eigen-values of the Laplacian on the 
cigar agrees with the expression (\ref{EV}) we have given in the 
previous subsection. In passing let us remark that we can obtain 
a minisuperspace analogue of the reflection amplitude (\ref{R}) 
from the corresponding quantity in \H\ (see e.g. \cite{teschnermini}) 
through  
$$ {\cal R}_0(j,n) \ = \ {\cal R}^{\rm H}(j,n,p=0) \ = \  
\frac{\Gamma(2j+1) \Gamma^2(-j+\frac{n}{2})}{ \Gamma(-2j-1) 
 \Gamma^2(j+1 - \frac{n}{2}) } \ \ . 
$$ 
This agrees with the results of the direct minisuperspace analysis 
in \cite{dvv} and it coincides with the semi-classical limit of the 
exact result (\ref{R}). 
\medskip 

Let us now describe the coset construction for the full conformal 
field theory. As usual, the primary field of the coset $\H/\U$ can 
be constructed in the product theory $\H \times \U$ of the $\H$ 
conformal field theory with a 1-dimensional free 
space-like
bosonic field. 
To be more precise, we continue by listing a few facts about the 
$\U$-model, i.e.\ the theory of a single free bosonic field 
$X$ that is compactified on a circle of radius $R = \sqrt {2k}$.
The field $X$ possesses the propagator 
$$ \langle X(z,\bar z) X(w,\bar w)\rangle \ = \ 
    \log |z - w|^2  
$$  
and satisfies $ X = X + 2 \pi R$. The latter condition instructs 
us to consider the following set of local exponentials,
$$ V_{n\, w}(z,\bz) \ :=\ e^{i \frac{2 \a}{\sqrt{2k}} X(z) + 
   i \frac{2\bar \a}{\sqrt{2k}} \bar X(\bz)} $$ 
where $\a = \frac12 (n + kw)$ and $\bar \a = \frac12(n - 
k w)$ and both $n$ and $w$ are integers. These exponential fields 
possess the following conformal dimensions
$$ h_{n\, w} \ = \ \frac{(n +kw)^2}{4k} \ \ \ \mbox{ and } 
\ \ \ \bar h_{n\, w} \ = \ \frac{(n-kw)^2}{4k} \ \ . $$     
Let us also introduce the two chiral $\U$ currents 
$$ J(z)\ =\ i \sqrt{\frac{k}{2}}\ \partial X(z,\bar z) \ \ \ 
\mbox{ and } \ \ \      
\bar J(\bar z)\ =\ i \sqrt{\frac{k}{2}}\ \bar \partial X(z,\bar z)
\ \ . $$ 
With respect to these currents, our vertex operators $V_{n,w}$ 
carry the charge $(\a,\bar \a)$.  
\smallskip

According to the usual rules of the coset construction, we can 
represent primary fields for the cigar as products of primary 
fields for the $\H$-model and vertex operators in the \U-theory. 
Using the same conventions as in \cite{pst}, we denote the 
primary fields of the $\H$ model by $\Phi^{\rm H,j}_{n,p}$. 
These fields carry a charge $(\frac12(n+ip),-\frac12(n-ip))$ with 
respect to the \H\ currents $(J^0,\bar J^0)$ (which are some 
components of the \SLC currents  of the \H\ model). Hence, we can 
build primary fields $\phi^j_{n\, w}$ which are uncharged under 
the currents $J^0+J$ and its conjugate $\bar J^0-\bar J$ by (see 
\cite{dvv}), 
\bea \label{phiphiV}
  \Phi^j_{n\, w} (z,\bz) \ = \  V_{n\, w}(z,\bz)  
     \, \Phi^{\rm H,j}_{n, -ikw} (z,\bz)  
\eea   
From this formula, the conformal dimensions of the coset field 
$\phi^j_{n\, w}$ can be easily determined, and the result  
agrees with eqs.\ (\ref{h}). 
\smallskip

To construct the characters of the coset model, we start from 
unspecialized characters of the \SL\ affine Lie algebra. For 
the continuous series of \SL\ representations in the $\H$ model, 
the latter read 
$$ \chi^{{\rm H},c}_{(j,\beta)}(q,z) \ = \ 
{\rm Tr}_{(j,\beta)}\,  q^{L_0 - \frac{c}{24}}\, z^{J^0_0} 
\ = \ \frac{q^{-\frac{(j+\frac12)^2}
{k-2}}}{\eta(q)^3} \ \sum_{r \in \Z} \ z^{r-\beta}\ \  
$$     
where $j \in -\frac12 + i \R$, $\beta \in [0,1[$ and $J_0^0$ denotes
the zero mode of the current $J^0$. We multiply these characters with 
the following unspecialized characters of the $\U$ model,   
\bea\label{charU} 
\zeta_\a(q,z) \ = \ \frac{z^\a \, q^{\frac{\a^2}{k}}}{\eta(q)}\ \ , 
\eea
to obtain characters of the product theory. Integration over the 
variable $\theta$ that appears in $z=\exp 2\pi i \theta $ 
along the entire real line gives the continuous 
characters $\chi^c_{(j,\a)}(q)$ of the coset model, 
\bea \label{ccharint} 
 \int_{-\infty}^\infty d\theta \ \chi^{\rm H,c}_{j,\beta}(q,z) 
   \zeta_\a (q,z) \ = \ \chi^c_{(j,\a)}(q) \ \eta(q)^{-2}\,  
  \sum_{r\in \Z} \ \delta(\alpha-\beta+r)\ \ . 
\eea    
Similarly, we can descend from the unspecialized characters of the 
discrete \SL\ representations 
\bea \label{Hcardisc}
\chi^{{\rm H},d}_j(q,z)\ =\ \frac{q^{-\frac{(j+\frac12)^2}{k-2}} \, 
  z^{-j-\half}}{{i \vt(q,z)}}\ \  
\eea 
to the characters (\ref{cardisc}) of the discrete series in the cigar,
\bea\label{dcharint} 
\int_{-\infty}^\infty d\theta \ \chi^{\rm H,d}_j(q,z) 
   \zeta_\a(q,z) \ = \ \chi^d_{(j,\a)}(q) \ \eta(q)^{-2}\,  
   \sum_{r \in \Z} \ \delta(j-\a + r)\ \ . 
\eea    
Actually, the integral over the real $\theta$-axis is singular. 
In the following we shall adopt a regularization in which we 
shift the contour by an amount $i\epsilon$. The computation 
of the integral is a bit more involved in this case. It can 
be found in Appendix A.  
\medskip 
 
{}From the construction of the fields $\Phi^j_{n\, w}$ we can 
immediately determine their correlation functions. In particular, 
since the compactified free boson $X$ does not contribute to the 
reflection amplitude, the cigar inherits its stringy reflection 
amplitude from $\H$, i.e.\ ${\cal R}(j,n,w) \ = \ {\cal R}^{\rm H}
(j,n,-i k w)$. An expression for ${\cal R}^{\rm H}$ can be found 
in \cite{moiii}.

\setcounter{equation}{0} 
\section{The D0-branes }

Our main aim is to discuss the exact solution for the point-like 
branes sitting at the tip of the cigar. We will start with a short 
semi-classical discussion in which we derive a minisuperspace formula 
for the coupling of closed strings to such  branes. Then we present 
exact expressions for the boundary states and the open string spectra
and we explain how they can be obtained by descent from $\H$. Finally, 
we show that our formulas are consistent with world-sheet duality. In 
this computation we shall see that the contributions from the discrete 
series of closed string modes are crucial. 

\subsection{Semi-classical description}

It is rather obvious from a Born-Infeld analysis why point-like 
branes are expected to sit at the tip of the cigar. In fact, for
a point-like brane, the Born-Infeld action reduces to  
\bea
S_{BI} \ \propto \  \int d^dy\, e^{-\Phi}\, \sqrt{\det (G + B + F)} \ 
   \propto \ \cosh \rho \ \ .
\eea
The string coupling takes its largest value at the tip so that the 
brane can minimize its mass at this point. Note that the semi-classical 
analysis does not 
provide any parameters for the D0-branes in the cigar. In the exact 
solution, on the other hand, we shall obtain a family of branes 
parametrized by one discrete parameter. 
\smallskip 

For comparison with the exact boundary state it is useful to spell 
out the semi-classical analogue of this quantity. In the minisuperspace
model we only have wave-functions $\phi^j_{n0}$ corresponding to closed 
string modes with vanishing winding number $w=0$. Their coupling to the
point-like brane at the tip of the cigar is simply given by the value of 
the function $\phi^j_{n0}$ at the point $\rho = 0$. Using the explicit 
expression (\ref{wavecig}) for the wave-function that we provided in the
previous section, we find 
\bea
\left(\langle \Phi^j_{n0}\rangle^{\rm D0}\right)_{k \rightarrow \infty} 
 \ = \  \phi^j_{n0}(\rho=0)\ =\ - \frac{\Gamma(-j)^2}{\Gamma(-2j-1)} 
   \ \delta_{n=0} \ \ . 
\label{1ptd0limit}
\eea
Note that due to the rotational symmetry of the point-like brane, only 
the modes with angular momentum $n=0$ have a non-vanishing coupling. This 
will remain true for the exact answer, but the coupling of the 
mode with $n=0$ will acquire a non-trivial dependence on the parameter 
$b=(k-2)^{-1/2}$ that controls the radius of the circle at infinity. 
\smallskip 

We can also predict some features of the open string spectrum of the
D0-brane. Since this brane is point-like, the spectrum should be
discrete. Moreover, the rotation symmetry of the D0-brane
allows the presence of states with arbitrary momentum $n$ in the open
string spectrum, and forbids the presence of winding open
strings. These expectations will be confirmed in the sequel.

\subsection{The exact solution}

Let us now summarize our results for the D0-branes in the cigar and 
compare them to semi-classical expectations. We claim that there exists 
a discrete family of localized branes living near the tip of the cigar, 
parametrized by some integer $m=1,2,\cdots$.\footnote{Within this 
section we allow $m$ to be any integer. There are some indications, 
however, that $m=1$ is the only value of $m$ that corresponds to a 
physical brane in the 2D black hole (see below).} For the one-point
functions of the bulk primary fields in the presence of these branes
we propose
\bea
\langle \Phi^j_{n w} (z,\bz) \rangle^{\rm D0}_{m} \  & = & \  
\delta_{n,0}\,  \N_m(b) \, (-1)^{mw} \, \left(\frac{k}{2}\right)^{\frac{1}{4}}
 \,  \frac{\Gamma(-j+\frac{kw}{2})
  \Gamma(-j-\frac{kw}{2})}{\Gamma (-2j-1)} \, \times \nonumber 
   \\[2mm]
 & & \hspace*{-1cm} \times \  
   \frac{\sin \pi b^2}{\sin\pi b^2 m} \ \frac{\sin \pi b^2 m (2j+1)}
     {\sin\pi b^2 (2j+1)}  
   \, \frac{\Gamma(1+b^2)\nu_b^{j+1}}{\Gamma(1-b^2(2j+1))} \  \
   \frac{1}{|z-\bz|^{h^j_{nw}+ \bar h^j_{nw}}}\ \ .    
\label{1ptd0}
\eea
In principle, there is some freedom in normalizing the one-point 
functions which gives rise to the numerical pre-factor $\N_m(b) = 
\la \Phi^{-1}_{00} (z=\halfi)\ra$. Here, we fix the value of this 
quantity to be 
\bea\label{Nd0} 
\N_m(b) \ = \ \left(\frac{b^2}{2}\right)^{\frac{1}{4}} \,
               \frac{\sin \pi b^2 m}{\sqrt{2\pi \sin \pi b^2}}\ \ .  
\eea
This choice is distinguished through our Cardy computation (see 
below) and it also arises when we descend from properly normalized 
boundary states in the \H model (see appendix B and next subsection). 
\smallskip 

There are several remarks we would like to make about our formula for 
the one-point function. To begin with, we should stress that it does not 
only give the couplings to states from the continuous series but that the 
expression is also meant to apply to the discrete spectrum of the theory.
For the latter, however, the quantity on the right hand side is singular. 
We anticipated such singularities before in our discussion of the reflection
amplitude (\ref{R}) for bulk fields. They were explained there as an  
artifact of our normalization. If we would renormalize fields from the
discrete series such that they have a finite two-point function, then 
their one-point functions would become finite as well.  

It is also worth pointing out that our formula shares many features with 
the related expression for point-like branes in Liouville theory 
\cite{zzb}. In our case, however, there is only a single discrete 
parameter $m$ instead of two, and all the associated branes possess 
a semi-classical limit $b \rightarrow 0$. If we set $m =1$ and restrict 
to states with zero winding number $w$, we recover the semi-classical 
prediction (\ref{1ptd0limit}). Brane couplings for $m\neq 1$ contain an 
additional factor $m$ when $b$ goes to zero. Hence, in the limiting regime 
they look like a collection of $m$ point-like branes at the tip of the cigar. 
\medskip 

The second result we want to state here concerns an expression for 
the spectrum of open strings stretching between two D0-branes with 
parameters $m$ and $m'$, respectively. We express this in the form 
of an annulus amplitude with modular parameter $q=\exp 2\pi i\tau$,
\bea
Z^\Dz_{mm'}(q) & = & \sum_{2J+1 = {{\rm max}(m,m')}}^{m+m'-1}\   
 \sum_{\ell\in \Z}\ \left[\chi
   ^d_{(J,\ell-J)}(q)-\chi^d_{(-J-1,\ell+J+1)}(q)\right] \ . 
\label{specd0}
\eea  
The representation labels $\ell-J$ and $\ell+J+1$ should be
interpreted as the momentum of the open strings in the angular
direction, whereas their winding is zero.
The latter fact can be traced back to the rotational symmetry 
of the point-like branes, 
% I rephrased the following sentences.
which prevents winding open strings from living on them (and also
implies that closed strings coupling to them must have zero angular
momentum).
Of course, even though discrete states with
zero winding do not appear in the closed string spectrum, this does
not prevent them from being physical in the open string theory. A
potentially more severe problem is the appearance of larges values of
$J$ for $m$ large enough,
since it threatens the unitarity of the spectrum. We take this as 
one indication that the large values of the parameter $m$ are 
unphysical. 
\smallskip

For two equal branes with the special labels $m = 1 = m'$, 
our formula (\ref{specd0}) for the partition function of 
D0-branes simplifies as follows, 
\bea
Z^\Dz_{11}(q) \ = \ \sum_{\ell \in \mathbb{Z}} \ 
  \left[ \chi^d_{(0,\ell)}(q) - \chi^d_{(-1,\ell-1)}(q)\right] \ 
\ = \ q^{-\frac{c}{24}}\, \left(1 + 2q^{1+\frac{1}{k}}
   +q^2 + O(q^{2+\frac{1}{k}}) \right)\ \ . 
\eea 
As for the point-like brane in Liouville theory, there appear no 
contributions from states with conformal weight $h=1$. This means 
that the point-like branes do not possess marginal deformations that 
can move them away from the tip of the cigar, in agreement with our 
geometric intuition. However, in the limit of large $k$, two states
become marginal. They correspond to displacements of a D0-brane
in the flat space limit of the 2-dimensional cigar. 
\smallskip 

Note also that at finite $k$ the identity field is the only relevant 
operator of the theory. The same is true for a stack of such branes 
and, unlike in many other backgrounds, their spectrum does not even contain 
marginally relevant fields. In a supersymmetric theory, the identity 
field is projected out and hence a stack of N D0-branes with parameter 
$m=1$ appears to be a stable state with D0-brane charge $N$. But our 
theory contains a candidate for another state with the same charge, 
namely the D0-brane with parameter $m=N$. At least in the semi-classical 
limit, the two D0-brane charges indeed agree. On the other hand, it is 
not hard to see that the D0-brane with parameter $m=N$ has lower mass 
than the stack of D0-branes with parameter $m=1$. Recall that  the mass 
of a D-brane is the one-point function of the identity operator. Hence, 
up to some $k$-dependent pre-factor, the mass of the $N^{th}$ D0-brane 
is given by $\sin \pi b^2 N$ which is strictly smaller than the 
total mass $N \sin \pi b^2$ of the stack. We can only reconcile 
this with the stability of the stack if we declare the D0-branes 
with $m > 1$ to be unphysical. Support for this solutions also 
comes from the analysis of the spectrum of open strings that 
stretch between D0 and D2-branes (see below). Let us finally also 
point out that even in a supersymmetric theory the D0-brane
spectra  with $m  > 1$ contain a large number of tachyonic modes 
and hence would be unstable.     

\subsection{D0-branes from descent}

We have to justify our expressions for the boundary states (\ref{1ptd0})
and spectrum (\ref{specd0}) of the D0-branes in the cigar. Our basic
strategy is to ``descend'' from the known \S-branes in \H\ \cite{pst}. 
The full consistency of those \S-branes has not been proved, but they 
satisfy two nontrivial consistency checks. First, their boundary state 
was shown to solve constraints coming from the factorization of bulk 
two-point functions \cite{pst,pon}. Furthermore, consistency with 
world-sheet duality has been checked. Since these branes are compact, 
the annulus amplitude is built from a discrete set of open string 
modes. A generalization of this Cardy-type computation is spelled 
out in Appendix B. 
\smallskip 

Let us point out that the \S-branes in \H\ do not have a clear 
geometrical interpretation. In fact, they look like spheres with an
\emph{imaginary} radius (see \cite{pst}). This is not too surprising
since \H also suffers from an imaginary Neveu-Schwarz $B$-field. On 
the other hand, such pathologies are cured when we pass down to the 
cigar and consequently the D0-branes
are good, physical objects.  
\medskip

To obtain the exact expression for the coupling of closed string 
modes to the D0-branes we follow the simple prescription that
is encoded in the product formula (\ref{phiphiV}) for the bulk 
field $\Phi^j_{n w}$. In other words, we construct the one-point 
functions for the cigar as a product of one-point functions for 
an \S-brane in \H with a one-point function of the \U\ vertex
operator $V_{n w}$. Following the usual ideology of brane 
constructions in coset models \cite{MMS,Gaw,FrSc}, we have to 
impose Neumann boundary conditions on the free boson. In fact, 
this choice guarantees that the resulting branes in the numerator 
theory extend in one direction, along the orbits of the \U\ 
action. After passing to the coset, we are left with a compact 
object again. 
\smallskip  

It is now easy to explain the origin of the formula (\ref{1ptd0}). 
To this end we rewrite the one-point function of the \S\ brane in 
\H\ (see \cite{pst}) in terms of the continuous momentum $p$ and
the angular momentum $n$,
\bea
\langle \Phi^{{\rm H},j}_{n p}(z,\bar z) \rangle_{m}^{S^2} & = & 
  \, \delta_{n,0} \, \N_m(b)\, \frac{\Gamma(-j+\frac{ip}{2})
  \Gamma(-j-\frac{ip}{2})}{\Gamma (-2j-1)} \, \times \nonumber 
   \\[2mm]
 & & \hspace*{-1cm} \times \  
   \frac{\sin \pi b^2}{\sin\pi b^2 m} \ \frac{\sin \pi b^2 m (2j+1)}
     {\sin\pi b^2 (2j+1)}  
   \, \frac{\Gamma(1+b^2)\nu_b^{j+1}}{\Gamma(1-b^2(2j+1))} \  \
   \frac{1}{|z-\bz|^{-\frac{2j(j+1)}{k}}}  \ \ .  
\label{1ptd0h3}
\eea
Here, $\N_m(b)$ is the same factor (\ref{Nd0}) as in the last 
subsection. Though this factor has not been spelled out in 
\cite{pst}, it is implicit in the discussion and we compute 
it explicitly in Appendix B.  
As explained above, the expression (\ref{1ptd0h3}) needs to be 
multiplied with the following one-point function for the \U\ 
vertex operators
\bea \label{1ptN} 
 \langle V_{n w} (z,\bar z) \rangle^N \ = \ 
 \delta_{n,0} \ \frac{(k/2)^{\frac{1}{4}}}{|z-\bar
   z|^{\frac{kw^2}{2}}}    
\eea
where the superscript $N$ stands for Neumann boundary conditions. 
If we finally replace $p$ by $-ikw$, we end up with the expression 
(\ref{1ptd0}) for the one-point function in the presence of D0-branes.%
\footnote{The extra factor $(-1)^{mw}$ in eq.\ (\ref{1ptd0}) has been 
introduced to ensure consistency with the world-sheet duality 
(see below).} 
\medskip 

In order to deduce our formula (\ref{specd0}) for the open string 
spectrum on the D0-branes, we start from the unspecialized partition 
function (\ref{specd0h3}) of the \S-branes. Each character in this
partition function can be decomposed into Fourier modes labelled by 
an integer $\ell$. This corresponds to the decomposition of affine 
\SL\ representations into sectors of the non-compact parafermions. 
Our prescription for the open-string spectrum on the D0-branes is 
simply to sum all those coset representations, just as in the 
construction of B-branes in $SU(2)/U(1)$ from maximally symmetric 
branes in $SU(2)$ \cite{MMS}.

\subsection{Cardy consistency condition  \label{cardyd0}}

We are now prepared to demonstrate that our exact solution  
is consistent with world-sheet duality, i.e.\ to show 
that after modular transformation the annulus amplitude 
(\ref{specd0}) can be re-interpreted as an overlap of the 
boundary states which are encoded in the one-point functions 
(\ref{1ptd0}). Though this computation can in principle be 
performed without any input from the $\H$ theory, using some 
of the quantities associated with the spherical branes on $\H$ 
turns out to be quite useful. The coset model computations, 
however, are more complicated because they involve contributions 
from discrete closed string states. There are no such discrete 
states in the bulk spectrum of \H\ and hence the overlap of the 
boundary states for spherical branes in \H\ contains no discrete 
contributions.On the cigar such terms are claimed to appear and 
we will see that this is indeed the case. 
\smallskip   

Our strategy here is to start on the open string side, i.e.\ with 
the amplitude (\ref{specd0}). The modular transformation to the 
closed string world-sheet is then performed in two steps. First 
we rewrite the open string amplitude on the cigar in terms of 
the corresponding quantity for spherical branes on \H\ using the
following formula, 
\bea \label{ZZS} 
Z^\Dz_{mm'}(q) \ = \ \eta^2(q) \int_{-\infty}^\infty d \theta 
   \int_{-\infty}^\infty d\a\, \zeta_\a(q,z) Z^{\S}_{mm'}(q,z) 
\ \ ,   
\eea 
where $\zeta_\a$ was defined in (\ref{charU}).
It follows directly from our description of the coset construction
(see formula (\ref{dcharint})) and can be verified by inserting the 
explicit expressions (\ref{specd0}) and (\ref{specd0h3}) for the 
two annulus amplitudes. In a second step, we modular transform the 
annulus amplitude of the spherical brane. Some relevant formulas 
can be found in Appendix B (see in particular eq.\ (\ref{intP})). 
After having rewritten the partition function in this way, we arrive 
at the expression
\bea \label{Zstep2} 
Z^\Dz_{mm'}(q)\ = \ -2 \eta(q)\, \sqrt\frac{b^2 k}{- i \tau}  
\int d\theta \int dP\ \frac{\tq ^{b^2P^2}}{\vt (\tq,\tz)} \sum_J 
\sinh 2\pi b^2 P (2J+1)\ \sinh 2\pi P \tth\ \ . 
\eea
Here, $q = \exp 2\pi i \tau$ and the summation over $J$ has the same 
range as in eq.\ (\ref{specd0}) so that, in particular, $(2J+1)$ is an 
integer. We also recall that the parameter $\tth$ appearing in $\tz = 
\exp 2\pi i \tth$ is given by $\tth = \theta/\tau$. In the derivation 
of eq.\ (\ref{Zstep2}) from eq.\ (\ref{ZZS}) we have performed the 
integral over $\s$.   
\smallskip 

Now we want to invert the integrations over $\theta$ and $P$. 
But the integrand has poles along the real $\theta$-axis. Hence, we 
first have to displace the $\theta$ integration contour slightly, away 
from the real axis to $\R+i\epsilon$. For definiteness we will assume 
that $\epsilon>0$, but the final answer does not depend on the sign of 
$\epsilon$. Using elementary techniques from the theory of complex 
functions it is possible to prove the following formula 
\bea
\int _{\R +i\epsilon} d\theta\ \frac{\sinh 2\pi P \tth}{\vt(\tq,\tz)}
 \ =\ \frac{\tanh \pi P}{\sqrt{-i\tau}\, \eta(q)^3} \ \sum_{n\in \Z} 
 \ (-1)^n \, \tq^{n^2/2-iPn} \ \ \ .
\eea
Insertion into our previous expression (\ref{Zstep2}) for the amplitude 
$Z^\Dz_{mm'}(q)$ gives 
\bea
Z^\Dz_{mm'}(q)\, =\, -2 \sqrt{b^2 k}  \ \int dP\ \sum_J 
  \sinh 2\pi b^2 P (2J+1)\ \frac{\tanh \pi P}{\eta(\tq)^2}
 \sum_{n\in \Z} \ (-1)^n\,  
\tq^{b^2(P - \frac{i}{2b^2} n)^2 + \frac{k}{4} n^2} \  .
\label{resim}
\eea
On the right hand side we have some series containing powers of 
the parameter $\tq$, but the exponents are complex. This prevents
a direct interpretation of these exponents as energies of closed
string states which would couple to the D-branes. To cure the
problem, we exchange the summation of $n$ with the integration 
over $P$ and we substitute $P$ by the new variable 
$$ P_n =  P - \frac{i}{2b^2} \, n \ \  $$
in each of the summands. $P_n$ is integrated along the line 
$\mathbb{R}-in/2b^2$. The crucial idea now is to shift all
these different integration contours back to the real line. 
This will give contributions associated with the continuous 
part of the boundary state. But while we shift the contours, 
we pick up residues from the singularities. The latter lead 
to terms associated with the discrete series. To work out the 
details, we split the partition function into a continuous and 
a discrete piece,   
$$ 
 Z^\Dz_{mm'}(q) \ = \ Z^c_{mm'}(q) + Z^d_{mm'}(q) \ \ . 
$$ 
Note that this split is defined with respect to the closed 
string modes. In terms of open string modes, our partition 
function contains only discrete contributions. According to 
our description above, the continuous part of the partition 
function reads as follows 
\bea
Z_{mm'}^{c}(q) \ = \ -2\sqrt {b^2k} 
\int dP\ \sum_{w \in \mathbb{Z}} 
\frac{\tq^{b^2 P^2+\frac{k}{4}w^2}}{\eta(\tq)^2} 
\ \sum_{J}\, (-1)^{2Jw}\, \frac{\sinh 2\pi b^2 P(2J+1)
  \sinh 2\pi P}{\cosh 2\pi P +\cos \pi kw }\  .
\label{specd0cont}
\eea
In writing this formula we have renamed the summation index 
from $n$ to $w$. It is now not difficult to see that this part 
of our partition function can be expressed as follows 
\bea
Z_{mm'}^{c}(q) & = &  
\int dP\ \sum_{w\in\Z} \ \chi^c_{(\j,\frac{kw}{2})}(\tq) \ 
  \Psi_m(\j,w) \, {\Psi}_{m'}(\j,w)^* \ \  
 \nonumber \ \ \mbox{ where } \\[2mm] 
 \Psi_m(j,w) & = &   (-1)^{mw} \, \left(\frac{kb^2}{\pi^2}
  \right)^{\frac{1}{4}} \ \frac{\Gamma(-j+\frac{kw}{2})
  \Gamma(-j-\frac{kw}{2})}{\Gamma (-2j-1)} \ \times 
   \\[2mm] & & \hspace*{4cm} \times \ 
   \frac{\sin \pi b^2 m (2j+1)}{\sin\pi b^2 (2j+1)}  
   \, \frac{\Gamma(1+b^2) \nu_b^{j+1} \sqrt{\sin \pi b^2}}
     {2\Gamma(1-b^2(2j+1))} \ \ .\nonumber 
\label{cyld0cont}
\eea
The coefficients $\Psi_m(\j,w)$ of the boundary state that 
we have introduced to present our result for $Z^d$ are easily 
recognized as the couplings of the one-point functions 
(\ref{1ptd0}).
\smallskip 

After this first success we now turn to the calculation of the 
discrete piece of the amplitude. It is clear from (\ref{resim}) 
that the residues we pick up while shifting the contours give the
following discrete contribution to the partition function  
\bea
\half Z^d_{mm'}(q) & =&  -2\sqrt{b^2k}\, 
 \sum_J 
\sum_{n=1}^\infty \ (-1)^n \tq^{\frac{k}{4}n^2}\, \times
\nonumber  \\[2mm] & & \hspace*{0cm} \times \,    
\sum_{m=0}^{E(\frac{k-2}{2}n-\half)}\  \sin \left(
\frac{2\pi (2J+1)(m+\half)}{k-2}\right)\
\frac{\tq^{-\frac{1}{k-2}\left(
    m+\half-\frac{k-2}{2}n \right)^2}}{\eta(\tq)^2}
\ \ .
\label{specd0disc}
\eea
Here, $E(x)$  denotes the integer part of $x$. A careful study of the 
energies which appear in this partition sum shows that the contributing
states can be mapped to discrete closed string states with zero momentum. 
The latter are parametrized by their winding number $w$ and by their spin 
$j$ or, equivalently, by the label $\ell=\frac{kw}{2}+j \in \Z$, and the 
level number $s$ as in the character (\ref{cardisc}). The map between the 
parameters $(w,\ell,s)$ and the summation indices $(m,n)$ of formula 
(\ref{specd0disc}) is
\bea
m\ =\ \ell-w \ \ \ \ , \ \ \ \ n\ =\ s+w\ \ .
\label{ident}
\eea
It can easily be inverted to compute the labels $(w,\ell,s)$ in terms
of $(m,n)$, 
\bea
w\ = \ E\left(\frac{2m+1}{k-2}\right) 
\ \ \ \ , \ \ \ \ \ell \ = \ m+E\left(\frac{2m+1}{k-2}\right) \ \
\ \ , \ \ \ \ s\ =\ n-E\left(\frac{2m+1}{k-2}\right)\ \ .
\label{identrecip}
\eea
In terms of $j$ and $w$, the partition function (\ref{specd0disc}) may 
now be rewritten as follows, 
\bea
 Z_{mm'}^{d}(q)\ = \ -2\sqrt{b^2k} \, \sum_J 
   \sum_{w\in \Z}\, \sum_{j \in \cJ^d_{0w}}\ 
(-1)^{2Jw}\, \sin\left(\frac{\pi(2J+1)(2j+1)}{k-2}\right)\, 
\chi_{(j,\frac{kw}{2})}(\tq)\ \ .
\label{cyld0disc2}
\eea
It remains to verify that the coefficients of the characters coincide 
with those derived from the boundary state (\ref{1ptd0}). In our 
normalization, the boundary coefficients $\Psi_m(j,w)$ are given 
through the same expression (\ref{cyld0cont}) as for the continuous 
series but we have to divide each term in the annulus amplitude by 
the non-trivial value of the bulk two-point function (\ref{twopt}) 
of discrete closed string states, i.e.\  
\bea
 {\cal Z}_{mm'}^{d}(q) & = &   
\sum_{w \in \Z} \, \sum_{j \in \cJ^d_{0w}}\ \frac{\Psi_m(j,w)
\Psi_{m'}(j,w)^*} {\la \Phi^{j}_{0w}\Phi^{j}_{0w}\ra} \  
\chi^d_{(j,\frac{kw}{2})}(\tq)\\[2mm]  
& = &  \sum_{w \in \Z} \, \sum_{j \in \cJ^d_{0w}}\ {\rm Res}_{x=j}
 \left(\frac{\Psi_m(x,w)\Psi_{m'}(x,w)^*}
  {{\cal R}(x,n=0,w)}\right) \ \chi^d_{(x,\frac{kw}{2})}(\tq)\ \ .
\label{cyld0disc}
\eea
The second line provides a more precise version of what we mean in 
the first line. Recall that the bulk two-point correlator contains 
a $\delta$-function which arises because of the infinite volume 
divergence. If we drop this $\delta$-function in the denominator, 
then the result has poles and the physical quantities are to read 
off from the residues. A short explicit computation 
shows that the argument of the Res-operation in eq.\ (\ref{cyld0disc}) 
indeed has simple poles at $ x \in \cJ^d_{0w}$ and that the residues
agree exactly with the coefficients in formula (\ref{cyld0disc2}), 
just as it is required by world-sheet duality.

\setcounter{equation}{0} 
\section{The D1-branes }

We now turn the construction of a 2-parameter family of 1-dimensional
branes on the cigar. All these branes stretch out to $\rho = \infty$
and hence they are non-compact. Our discussion follows the pattern of 
the previous section. We shall begin with some remarks on the 
semi-classics and then spell out the one-point function and the open 
string spectrum. Because of the non-compactness of the branes' 
world-volumes, their open string spectrum is continuous and it 
involves some non-trivial spectral density. Both quantities, the 
exact one-point functions and the open string spectra, are obtained 
rather easily from the corresponding quantities for the Euclidean 
\AA branes in \H (see \cite{pst,deuxbranes}) since there are no 
complications associated with the discrete series on the cigar. The 
section concludes with a Cardy-type consistency check of world-sheet
duality.

\subsection{Semi-classical description} 

The D1-branes in the cigar are most easily studied using a new coordinate 
$u=\sinh\rho$ along with the usual angle $\theta$. We have $u\geq 0$ and 
$u=0$ corresponds to the point at the tip of the cigar. In the new 
coordinate system, the background fields read 
\bea
ds^2 \ =\ \frac{k}{2}\, \frac{du^2+u^2d\theta^2}{1+u^2}\ \ \ \ , 
 \ \ \ \ 
e^\Phi\ =\ \frac{e^{\Phi_0}}{(1+u^2)^{\frac{1}{2}}}\ \ .  
\eea
When we insert these background data into the Born-Infeld action for 
1-dimensional branes we obtain 
\bea 
S_{\rm BI} \ \propto \ \int dy\, \sqrt{u'{}^2+u^2\theta'{}^2}\ \ ,
\eea
where we used the same conventions as in section 3.1 and the primes denote 
derivatives with respect to the world-volume coordinate $y$ on the D1-brane. 
It is now easy to read off that D1-branes are straight lines in the plane 
$(u=\sinh\rho,\theta)$. These are parametrized by two quantities, one being 
their slope, the other the transverse distance from the origin. In our 
original coordinates $\rho,\theta$, this 2-parameter family of 1-dimensional 
branes is characterized by the equations 
\bea
\sinh\rho\ \sin(\theta-\theta_0) \ =\ \sinh r \  .
\label{eqd1}
\eea
Note that the brane passes through the tip if we fix the parameter $r$ to 
$r=0$. All branes reach the circle at infinity ($\rho= \infty$) at two 
opposite points. The positions $\theta_0$ and $\theta_0+\pi$ of the latter 
depend on the second parameter $\theta_0$ (see figure \ref{70cigd1}).
\medskip 

\Fig{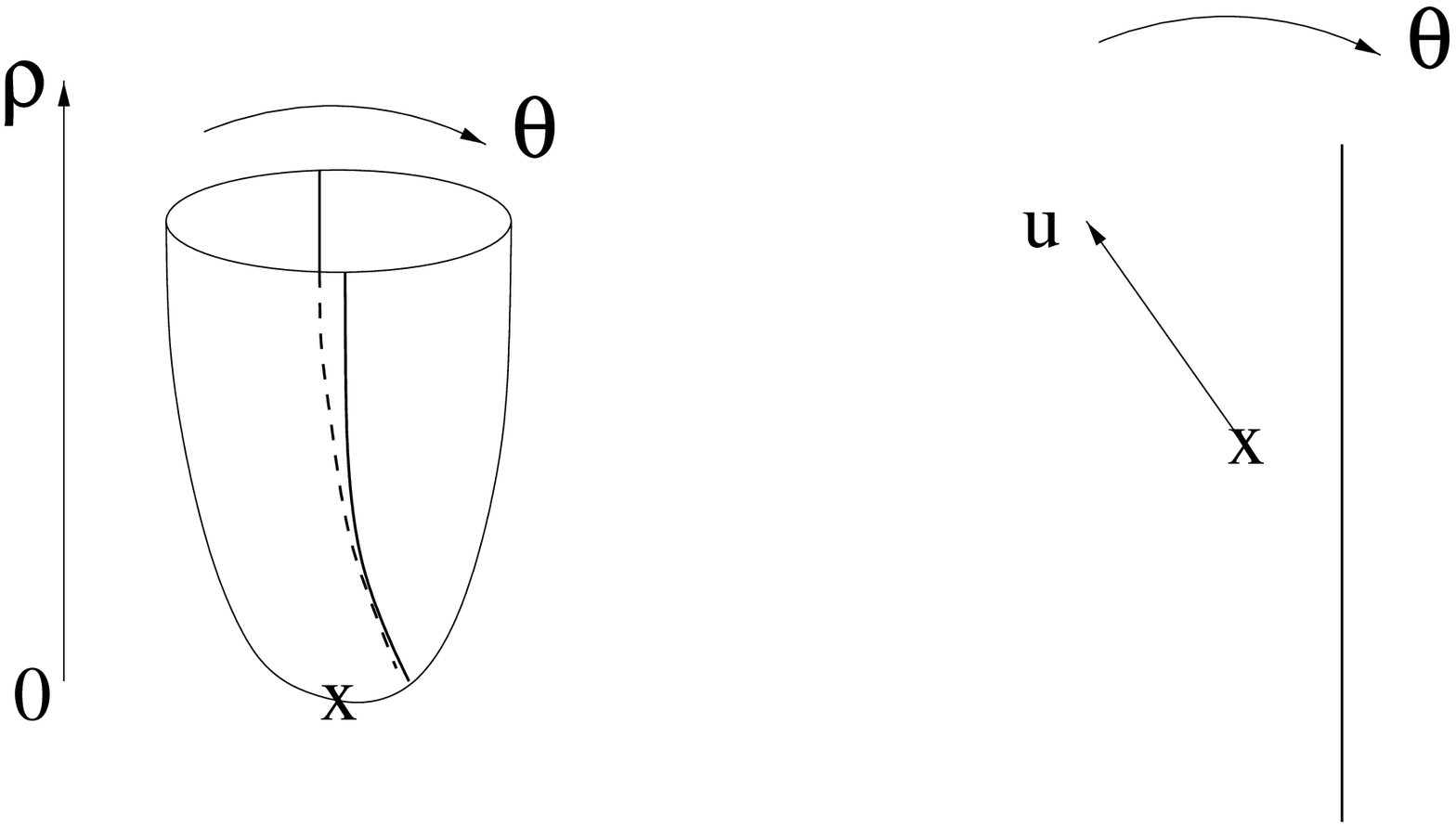}{10}{A D1-brane in the cigar in $\rho$ and $u$ coordinates}

Now that we have some idea about the surfaces along which our branes
are localized we can calculate their coupling to closed string modes
in the semi-classical limit. This is done by integrating the 
minisuperspace wave functions (\ref{wavecig}) of closed string modes
over the 1-dimensional surfaces (\ref{eqd1}). The result of this 
straightforward computations is the prediction 
\bea
\left(\langle \Phi^j_{n0}\rangle^{\rm D1}_r\right)\pp \ = \ 
e^{ i n\theta_0}\, \frac{\Gamma(2j+1)}{\Gamma(1+j+\frac{n}{2})
\Gamma(1+j-\frac{n}{2})}
\left( e^{-r(2j+1)}+(-1)^n e^{r(2j+1)}\right)\ .
\label{1ptd1limit}
\eea
The minisuperspace model of the cigar does not include any states
associated with closed string modes of non-vanishing winding number. 
But in the case of the D1-branes, experience from the analysis of 
branes on a 1-dimensional infinite cylinder teaches us that closed 
string modes with $w\neq 0$ do not couple at all. Since the 
discrete closed string modes only appear at $w \neq 0$, they 
are like-wise not expected to couple to the D1-branes. Consequently,
our formula (\ref{1ptd1limit}) predicts the semi-classical limit of
the only non-vanishing couplings in the theory.  
\medskip 

Let us now come to the semi-classical analysis of the open string 
spectrum. Because of the non-compactness of the 1-dimensional branes, 
this is a much more interesting topic than for the D0-branes. From 
the geometry we have just described, it is rather obvious that the 
spectrum of open strings on the D1-branes cannot contain any momentum 
modes along the $\theta$ direction. Open string winding modes around the 
compact circle, however, will appear in the exact spectrum and since there
are two branches of our branes at large $\rho$, the winding number is 
expected to be a half-integer number (see figure \ref{73cigcorde}). 
Note that this winding number is 
certainly not conserved in physical processes since open strings can 
`unwind' in the interior where the two branches of our 1-dimensional
branes come together.     
\smallskip 

\Fig{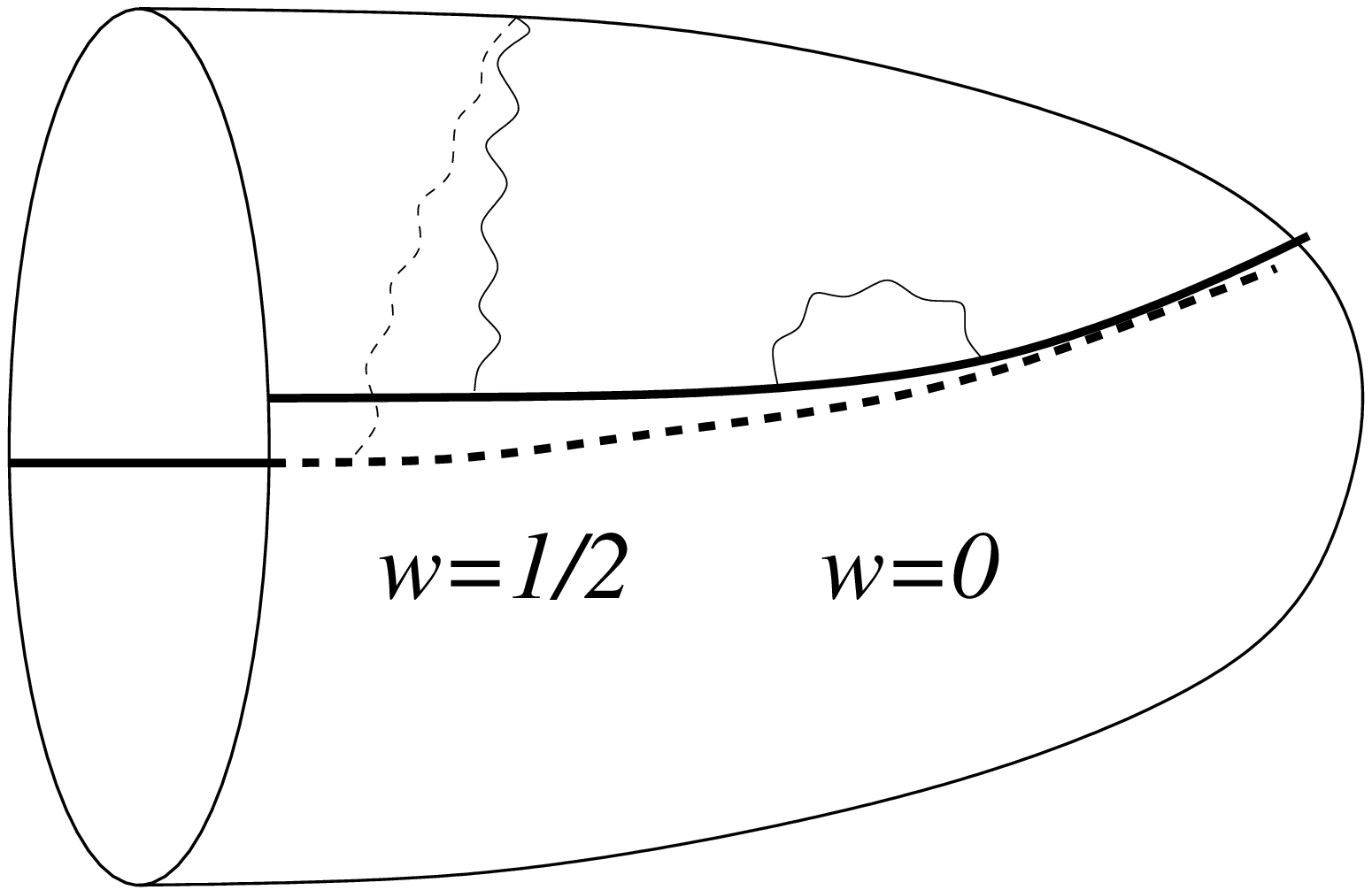}{7}{A D1-brane in the cigar with open strings on it }

It is also clear from the geometry of the D1-branes that open strings 
can be sent in from infinity with any non-negative real momentum $P$ 
along the radial direction. In other words, open strings on the 
1-dimensional branes possess a continuous spectrum. The latter can 
be characterized by an interesting new quantity: an open string spectral 
density. As we have argued before, the minisuperspace model cannot say 
anything on strings with non-zero winding number. Open strings with zero 
winding, however, possess a semi-classical point-particle limit. Hence, 
for the modes with $w=0$, we can make some prediction for the density 
$N(j,w=0|r,r)$ of radial open string momenta $P =-i(2j+1)$. In fact, by 
very general arguments (see e.g. \cite{pst}), the quantity $\rho$ 
is related the so-called reflection amplitude $R(j)$ by\footnote{A 
more careful discussion must take into account that the spectral 
density  itself is divergent. The divergence, however, is 
universal, and it can be removed, e.g.\ by considering relative 
spectral densities.} 
\bea \SN(\j) \ \sim \ \frac{1}{2\pi i} \, 
     \frac{\partial}{\partial P} R(\j) \ \ . 
\label{rhoR}\eea    
By definition, the reflection amplitude is the ratio between the 
coefficients of the in-coming an out-going plane waves within a 
wave function. For the minisuperspace limit of open strings on D1
branes we can compute $R(j)$ either through a direct study of 
the 1-dimensional Laplacian on the D1-branes or by descent from 
the corresponding minisuperspace model of open string on 
Euclidean \AA-branes. Both ways lead to 
the same answer,   
\bea
R(P,w=0|r,r)\pp \ =\ - (\cosh r)^{2iP}\frac{\Gamma(1-iP)\Gamma(\half+iP)}
{\Gamma(1+iP)\Gamma(\half-iP)}\ \ .
\label{refd1limit}
\eea 
Because of the rotational symmetry, the reflection amplitude does 
not depend on the angular parameter $\theta_0$. The geometrical 
picture we have outlined in this subsection along with the two 
precise predictions (\ref{1ptd1limit}),(\ref{rhoR},\ref{refd1limit}) 
we made for the semi-classical limit of the bulk one-point functions 
and the open string spectral density will be confirmed nicely through 
our exact solution.

\subsection{The exact solution}

In this subsection we shall present our proposal for exact boundary 
conformal field theory quantities describing 1-dimensional branes 
on the cigar. We claim that the exact solution is parametrized by 
two continuous parameters $r$ and $\theta_0$, just as in the 
semi-classical limit. Expressions for the one-point functions of 
closed string fields $\Phi$ in the D1-brane background were already 
proposed in \cite{pst},   
\bea
\langle \Phi^j_{nw} (z,\bz)\rangle^{\rm D1}_{(r,\theta_0)} & = & 
  \, \delta_{w,0} \,\N'(b)\, e^{ i n\theta_0} \, (2k)^{-\frac{1}{4}}\  
\frac{\Gamma(2j+1)}{\Gamma(1+j+\frac{n}{2})\Gamma(1+j-\frac{n}{2})} 
\, \times \nonumber \\[2mm]
& & \hspace*{-2cm} \times \, \left(e^{-r(2j+1)}+(-1)^n e^{r(2j+1)}
\right)
\, \Gamma(1+b^2(2j+1))\,\nu_b^{j+\half} \ 
\frac{1}{|z-\bar z|^{h^j_{nw}+\bar h^j_{nw}}}  \ \ 
\label{1ptd1}
\eea
where the prefactor is $\N'(b) = (8b^2)^{-1/4}$. 
Of course, the expression (\ref{1ptd1}) correctly reproduces the 
semi-classical expression (\ref{1ptd1limit}). The precise matching 
justifies the identification of the parameter $r$ and $\theta_0$ 
in the one-point function with the 
geometrical parameters $r$ and $\theta_0$ from the previous subsection.  
We also point out that the one-point functions possess poles whenever 
$b^2 (2j+1) = i b^2 P$ is a negative integer. As in the case of 
extended branes in Liouville theory \cite{zzf}, these poles can 
be traced back to the non-compactness of our branes. 
\medskip 

In addition, we want to spell out the density of open strings living on a 
D1-brane. For symmetry reasons this depends only on the parameter 
$r$, not on the angle $\theta_0$. 
As usual, the large volume divergence of the spectral density 
can be regularized by introducing a cut-off in the coordinate $\rho$. 
Since the divergence of the spectral density is universal, the 
cut-off can be sent to infinity in relative spectral densities. 
To be more specific, we will subtract the spectral density of 
the brane with label $r=0$, i.e.\ consider the quantity 
\bea
\Delta \SN(P,w|r,r)\ :=\ \frac{1}{2\pi i}\, \frac{\p}{\p P}
\log\frac{R(-\half+iP,w|r,r)}{R(-\half+iP,w|0,0)}\ \ . 
\label{reflection}
\eea
In this formula we have expressed the relative spectral density 
through a quotient of the associated stringy reflection amplitude.
We can now express our proposal for the relative spectral density 
in terms of formulas for $R(\j,w|r,r)$. As we observed previously,  
the open string winding number $w$ is expected to take values in  
$w \in \frac12 \mathbb{Z}$. Our formulas for $R(j,w)$ will turn 
out to depend on whether $w$ is integer or not. In the former case, 
$R(j,w|r,r)$  is given by 
\bea
R(\j,w\in\Z|r,r) & = & \nu_b^{iP}\, \frac{\Gamma ^2_k(\half-iP+\frac{1}{b^2})}
     {\Gamma ^2_k(\half+iP+\frac{1}{b^2})} \ 
   \frac{\Gamma_k(2iP+\frac{1}{b^2})}{\Gamma_k(-2iP+\frac{1}{b^2})}
   \ \frac{S^{(i)}_k(\rr+P)}{S^{(i)}_k(\rr-P)}\ \ 
\label{solrr} \\[2mm] 
\mbox{with} \ \ \ 
\log S^{(i)}_k(x) & = & \log S^{(0)}_k(x)\ =\ 
  i\int_0^\infty\frac{dt}{t}\, \left(\frac{\sin
    2tb^2x}{2\sinh b^2t\sinh t}-\frac{x}{t}\right), 
\label{fctsk}
\eea
Note that the reflection amplitude does 
not depend on the open string winding number $w$ as long as it is 
integer. For reasons that we will review below, open strings with 
integer winding number on a D1-brane in the cigar possess the same 
spectral density as open strings on an \AA\ brane in \AAA\ \cite{pst}. 
Similarly, for open string modes with $w \in \frac12 + \mathbb{Z}$
we propose a reflection amplitude of the form (\ref{solrr}) but with 
\bea 
\log S^{(i)}_k(x) \ = \ \log S^{(1)}_k(x) \ =\ 
  i\int_0^\infty\frac{dt}{t}\, \left(\frac{\cosh t\sin
    2tb^2x}{2\sinh b^2t\sinh t}-\frac{x}{t}\right)\ \ . 
\label{fctspk}
\eea 
The associated spectral density coincides with the spectral density
of open strings stretched between two opposite \AA\ branes in \AAA\ 
\cite{deuxbranes}. For this reason we shall often denote $\Delta
\SN(P,w),w \in \frac12 + \mathbb{Z},$ by $\Delta\SN(P|r,-r)$, and 
we shall also use the symbol $R(P|r,-r)$ for the reflection amplitude 
$R(P,w|r,r)$ with $w \in \frac12+\mathbb{Z}$.  
\smallskip 

Finally, we provide a prescription for how to compute the 
density of open strings stretching between two D1-branes with
different labels $r,r'$. At the same time we want both branes 
to possess identical angular parameter $\theta_0$. This implies
that the two branes coincide at $\rho = \infty$ so that the 
winding number $w$ continues to take values in $\frac12 
\mathbb{Z}$. The density of open strings stretching 
between two such branes $(r,\theta_0)$ and $(r',\theta_0)$ 
is given by (see \cite{deuxbranes} for a related formula in 
the case of \AA branes in \AAA) 
\bea
\Delta \SN(P,w|r,r')\ =\ \Delta \SN\left(P,w\left|\frac{r+r'}{2},
 \frac{r+r'}{2}\right.\right)+\Delta \SN\left(P,w\left|\frac{r-r'}{2},
\frac{r'-r}{2}\right.\right) \ \ . 
\label{specd1int}
\eea
This relative spectral density features in the following expression 
for the relative open string partition function of the D1-branes, 
\bea \left(Z^\Do_{(r,\theta_0)(r',\theta_0)} - Z^\Do_{(0,\theta_0)
 (0,\theta_0)}\right)(q)  \ = \  \int dP \, \sum_{w \in \frac12 \Z} 
    \ \Delta \SN(P,w|r,r')  \ \chi^c_{(\j,w)}(q) \ \ . 
\label{pfd1} \eea
In the special case $r=r'$ we shall later reproduce quantity
(\ref{pfd1}) through world-sheet duality starting from the one-point 
functions for D1-branes that we have proposed above.  
\smallskip 

Let us also briefly point out that for $r = r'$ and $w=0$, the 
semi-classical limit of our expression for the relative spectral 
density $\Delta \SN$ agrees with the expression (\ref{refd1limit}) 
above. In fact, with the choices we have made it is easy to show that  
\bea
\lim _{k\rar \infty} \Delta\SN(P,0|r,r)\ =\ \frac{1}{\pi}
\int_0^\infty \frac{dt}{t}\left( \frac{\cos\frac{2tr}{\pi}}
{\sinh t}-\frac{1}{t}\right) \ = \ \frac{1}{\pi}\log\cosh r \ \ . 
\eea
We would like to stress that the quantity $\Delta \SN (r,-r)$ 
which appears in $\Delta \SN(r,r')$ does not possess a well 
defined semi-classical limit. Given the geometrical setup we 
have drawn above, this does not come as a big surprise, but 
it still can be considered a test of the conformal field theory 
solution.  

%SR : added a paragraph

For completeness, let us mention that we also expect discrete open
string states to live on D1-branes in the cigar. This is a consequence
of similar results for $AdS_2$ branes in $AdS_3$ (reviewed in 
\cite{RibPhD}). However, this discrete part of the spectrum 
does not depend on the brane's parameter and thus does not contribute to
relative spectral densities.

\subsection{D1-branes from descent}

To obtain the D1-branes on the cigar from the \H\ model, we follow 
the same strategy as in subsection 3.3, only that this time we start 
with the Euclidean \AA\  brane on \H. Furthermore, when we insert 
the product formula (\ref{phiphiV}) into the one-point function, we 
have to impose Dirichlet boundary conditions on the free boson. 
With such a choice, we obtain a 2-dimensional brane in the 
numerator theory which then descends to a D1-brane on the cigar. 
\smallskip 

We recall from \cite{pst} that the Euclidean \AA-branes are 
characterized by the following set of one-point functions    
\bea
\langle \Phi^{{\rm H},j}_{np} (z,\bz)\rangle^{\AA }_{r} & = & 
  \delta(p) \, \N'(b) \  
\frac{\Gamma(2j+1)}{\Gamma(1+j+\frac{n}{2})\Gamma(1+j-\frac{n}{2})} 
\, \times \nonumber \\[2mm]
& & \hspace*{-2cm} \times \, \left(e^{-r(2j+1)}+(-1)^n e^{r(2j+1)}\right)
\, \Gamma(1+b^2(2j+1))\,\nu_b^{j+\half} \ 
\frac{1}{|z-\bar z|^{2\Delta^{\rm H}_j}}  \ \ .
\label{1ptads2}
\eea
The pre-factor $\N'(b)$ is the same as in eq.\ (\ref{1ptd1})
We also need the one point functions of the \U vertex operators. 
In case of Dirichlet boundary conditions on the free boson they
read 
\bea \label{1ptD}
\langle V_{nw} (z,\bar z) \rangle^D \ = \ 
 \delta_{w,0} \ \frac{(2k)^{-\frac{1}{4}}}{|z-\bar
   z|^{\frac{n^2}{2k}}} 
  \ \ ,   
\eea
Our expression (\ref{1ptd1}) corresponds to taking the product 
of eqs.\ (\ref{1ptads2}) and (\ref{1ptD}) with $p$ being replaced 
by $-ikw$. 
\smallskip 

The \AA\ branes in \H\ extend along the time direction which we
Wick-rotate and then gauge in order to obtain the cigar. Thus, in 
the terminology of \cite{MMS}, the D1-branes in the cigar should be 
thought of as non-compact A-branes (see figure \ref{74cigd1ads}). As 
in the compact case, the properties of A-branes are very easily 
deduced from those of parent branes. In particular, the open string 
density of states on a single brane is given by the same function 
$N(P|r,r)$. 

\Fig{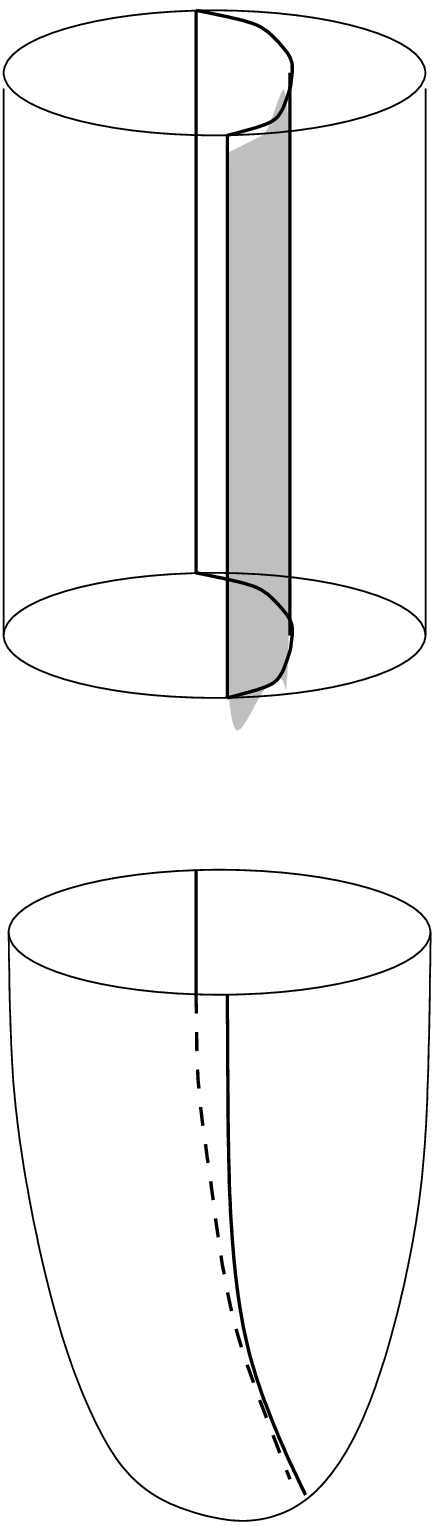}{3}{A D1-brane in the cigar descending from an \AA\
  brane in \H }

Only the appearance of the winding number $w$ in the open string 
spectrum makes the story a bit more subtle. In the context of \AA\
branes in \AAA, we interpret $2w$ as counting the number of spectral 
flow operations. It is well known that for $r \neq 0$ only an even 
number of spectral flow operations preserves the spectrum of open 
strings living on an \AA\ D-brane \cite{pr,plot}. If we apply an 
odd number of spectral flow operations to an open string with two 
ends on the same \AA\ brane then one of its ends gets mapped to the
opposite D-brane with parameter $-r$, see figure \ref{77cigflot}. 
These considerations explain why the spectral density for D1-branes
on the cigar depends only on the parity of $2w$ and they identify 
the density of open strings with $w\in \Z+\half$ with the quantity 
$N(P|r,-r)$ that was computed for branes in \H (see \cite{deuxbranes}). 
The same reasoning applies to the more complicated case of open 
strings stretched between two D1-branes with different parameters
$r,r'$. 

\Fig{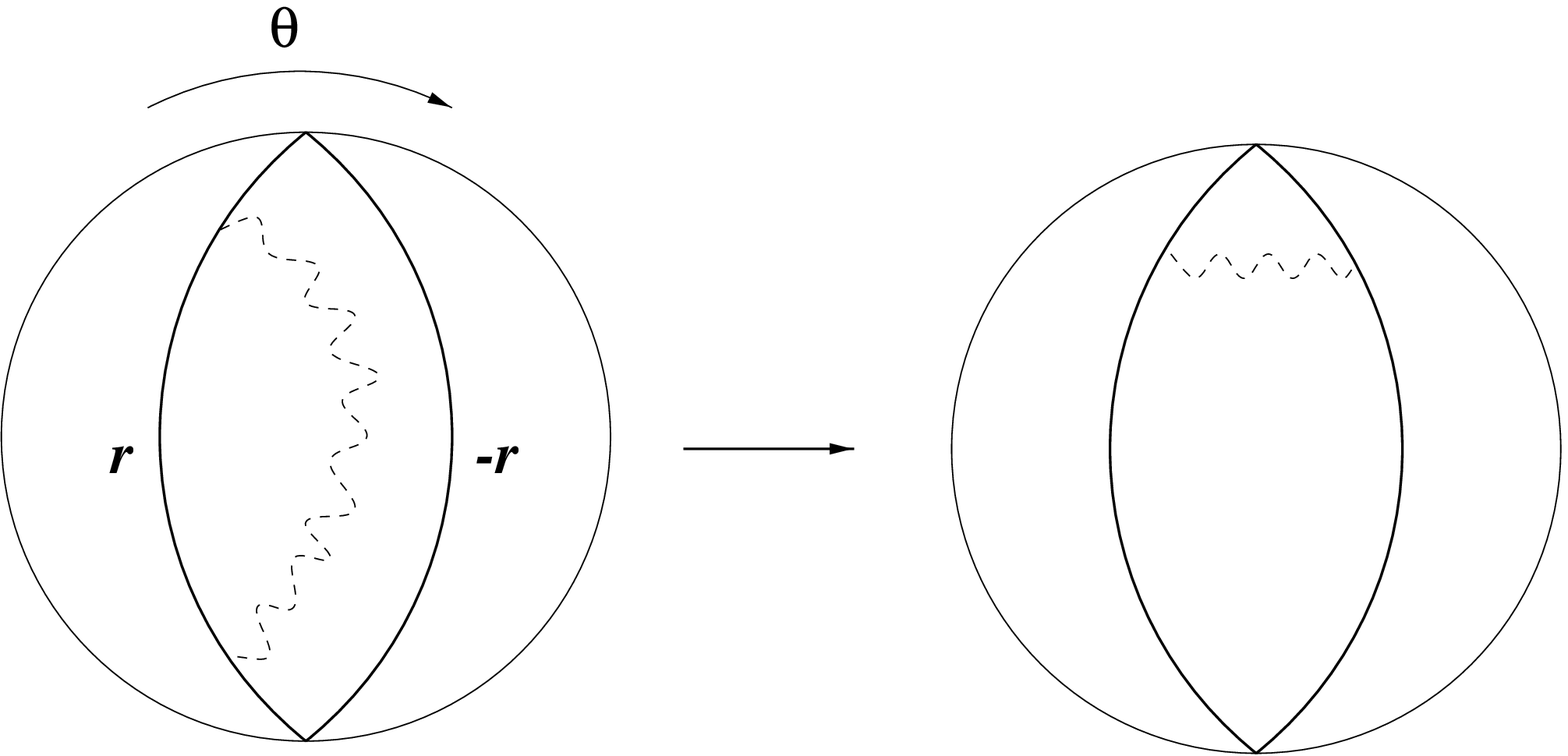}{8}{In \AAA\ viewed from above, the action of spectral
  flow on an open string ending on an \AA\ brane }

\subsection{Cardy consistency condition}

As in our discussion of the D0-branes we conclude the section in the
D1-branes by showing that the exact solution we have proposed is 
consistent with world-sheet duality. Our presentation will be rather 
brief since most of it is rather similar to the case of \AA branes
in \AAA (see \cite{pst,deuxbranes}). In particular, we will restrict 
our computations to the case $r = r'$.  
\smallskip 

Our main aim is to show how the two contributions from $w \in \mathbb{Z}$ 
and $w \in \frac12 + \mathbb{Z}$ in the open string partition function 
(\ref{pfd1}) arise from world-sheet duality. To this end, we start 
from the closed string amplitude\footnote{To be precise, this quantity 
again diverges with the volume and to make it well defined, one should 
either divide by the volume divergence or subtract the same amplitude 
for a fixed reference brane $r=0$ (see the discussion in \cite{pst}).}     
\bea
Z_{(r,\theta_0)(r',\theta_0)}(\tq) & = & 
\int dP\ \sum_{n\in\Z}  \ \chi^c_{(P,n)}(\tq) \ \Psi_{(r,\theta_0)}(\j,n) 
 \, \Psi_{(r',\theta_0)}(\j,n)^\ast  \ \ \ \mbox{where} \nonumber \\[2mm]      
\Psi_{(r,\theta_0)}(j,n) & = &  
 (kb^2)^{-\frac{1}{4}} \, e^{ i n\theta_0} \ 
  \frac{2 \Gamma(2j+1)}{\Gamma(1+j+\frac{n}{2})
 \Gamma(1+j-\frac{n}{2})} \, \times \\[2mm] 
& & \hspace*{2cm} \times \, \left(e^{-r(2j+1)}+(-1)^n e^{r(2j+1)}\right)
\, \Gamma(1+b^2(2j+1))\,\nu_b^{j+\half}\ \  \nonumber  
\eea
are the couplings of closed strings to D1-branes which can be read off 
from eq.\ (\ref{1ptd1}). In our computation of this amplitude for the 
special case $r=r'$ we shall focus on the $n$-dependence since this is 
what distinguishes the calculation from the case of \AA\ branes in 
\H. For the integrand of the amplitude $Z$ we obtain 
\bea
\lefteqn{ \sum_{n\in\Z} \, \tq^\frac{n^2}{4k}\Psi_{(r,\theta_0)}(\j,n) 
 \, \Psi_{(r',\theta_0)}(\j,n)^\ast  \ \propto } \hspace{1.5cm}
  \nonumber \\[2mm] 
& \propto &  \cosh ^2\pi P\ \cos ^22rP \sum _{n\in2\Z}
 \tq ^\frac{n^2}{4k} + \sinh^2\pi P\ \sin ^2 2rP 
 \sum _{n\in 2\Z+1}\tq^\frac{n^2}{4k}\ \ . \nonumber 
\eea 
Here we used that the quantity $\Psi(.,n) \Psi(.,n)^\ast$ depends on 
$n$ only through its parity. Nevertheless, the infinite sum over $n$ 
is convergent thanks to $n$-dependent factor $\exp(-2\pi i n^2/4k\tau)$ 
that comes from the coset character (\ref{cardisc}). It is now 
straightforward to rewrite the previous quantity as 
\bea
 & = & (\cosh ^2\pi P\ \cos ^22rP + \sinh
  ^2\pi P\ \sin ^2 2rP)\ \frac{1}{2}\, \sum _{n\in \Z} \ \tq
    ^\frac{n^2}{4k}+  
\\[2mm] 
&& \hspace*{5mm} + (\cosh ^2\pi P\ \cos
   ^22rP - \sinh 
   ^2\pi P\ \sin ^2 2rP)\ \frac{1}{2}\, (\sum _{n\in 2\Z}-\sum _{n\in
     2\Z+1}) \ \tq ^\frac{n^2}{4k}\ 
\\[2mm]
&\propto&
\cos ^22rP\ \sum_{w\in \Z}\ q^{kw^2}\ +\ \cosh 2\pi P\ \cos^22rP\
\sum_{w\in \Z+\half}\ q^{kw^2}\ +\ r-{\rm indep.}
\eea
The last step was a Poisson resummation which changed $n$ into $w$. 
This shows how the two sectors with $w\in\Z$ and $w\in \Z+\half$ 
appear in the open-string spectrum and that their densities differ
by a factor $\cosh 2\pi P$. After modular transformation, the extra
factor gives rise to the $\cosh t$ factor that we introduced in the 
definition of the special function (\ref{fctspk}) and that is not
present in the corresponding quantity (\ref{fctsk}) for integer 
winding number $w$. 

\setcounter{equation}{0} 
\section{The D2-branes }

After having discussed D0 and D1-branes we are left with one more 
species of branes. These are 2-dimensional objects that extend
all the way to $\rho = \infty$. They can carry a 2-form gauge 
field $F$ and hence are characterized by one real parameter. In 
the first subsection we shall argue for the existence of such 
branes using the Born-Infeld action. The explicit formulas we 
derive for the various open string background fields can be 
employed to derive a minisuperspace formula for the spectral 
density of open string states ending on the D2-branes. We 
present our exact solution 
for D2-branes in the second subsection before explaining how 
it may be obtained by descending from certain 2-dimensional 
branes in $\H$. Finally, we check the consistency of our 
proposal for the bulk one-point function and the open 
string spectral density with world-sheet duality. We shall 
show, in particular, how discrete open string modes on the 
D2-brane emerge within this Cardy computation.   

\subsection{Semi-classical description}

The main feature that distinguishes the D2-branes from the branes 
we have discussed above is that they can carry a world-volume 2-form 
gauge field $F = F_{\rho\theta} d\rho \wedge d\theta$. In the presence 
of the latter, the Born-Infeld action for a D2-brane on the cigar becomes  
\bea
S_{\rm BI} \ \propto \ \int d\rho d \theta \cosh \rho 
  \sqrt{\tanh^2\rho + F_{\rho\theta}^2}\ \ . 
\eea
We shall choose a gauge in which the component $A_\rho$ of the gauge 
field vanishes so that we can write $F_{\rho \theta} = \partial_\rho 
A_\theta$. A short computation shows that the equation of motion for 
the one-form gauge potential $A$ is equivalent to  
\bea
F_{\rho\theta}^2 \ = \ \frac{\beta ^2\tanh^2\rho}{\cosh^2\rho-\beta
    ^2}\ \ . 
\label{fd2}
\eea
If the integration constant $\beta$ is greater than one, then the
D2-brane is localized in the region $\cosh\rho\geq \beta$, i.e.\ 
it does not reach the tip of the cigar. We will exclude this case 
in our semi-classical discussion and assume that $\beta =\sin \s 
\leq 1$. The corresponding D2-brane cover the whole cigar. 
Integrating eq.\ (\ref{fd2}) for the field strength $F_{\rho \theta} 
= \partial_\rho A_\theta$ furnishes the following expression for 
the gauge potential    
\bea
 A_\theta(\rho)\ =\ \s-\arctan\left(\frac{\tan
    \s}{\sqrt{1+\frac{\sinh^2\rho}{\cos^2 \s}}}\right)\ \ .
\eea
In our normalization $A_\theta(\rho =0) = 0$, the parameter $\s$ 
is the value of the gauge potential $A_\theta$ at infinity. When 
this parameter $\sigma$ tends to $\sigma = \halfpi$, the F-field 
on the brane blows up. We should thus consider $\sigma = \halfpi$ 
as a physical bound for $\sigma$. 
\smallskip 

Let us point out that the F-field we found here vanishes at $\rho = 
\infty$. In other words, it is concentrated near the tip of the cigar. 
By the usual arguments, the presence of a non-vanishing F-field implies 
that our D2-branes carry a D0-brane charge which is given by the 
integral of the F-field. Like the F-field itself, the D0-brane 
charge is localized near the tip of the cigar, i.e.\ in a compact 
subset of the 2-dimensional background. Hence, one expects the D0-brane 
charge, and therefore the parameter $\sigma$ of the D2-brane, to be 
quantized. Another way to argue for such a quantization is through 
semi-classical charge conservation \cite{bfpr}. Note that the 
difference of two D2-branes with parameters $\sigma$ and $\sigma'$ 
is a 2-sphere if we do not allow for deformations of the circle at 
infinity. This implies that the difference of the D2-brane parameters
must be an integer $m$,    
\cite{bfpr}
\bea
\s-\s'\ =\ 2\pi \frac{m}{k}\ \ , \ \ m\in \Z\ \ . 
\label{qd2clas}
\eea
We shall see later that a `quantum corrected' version of this condition 
is needed in order to obtain a sensible spectrum of open strings between 
the two D2-branes.
\medskip 
 
Our next aim here is to predict the semi-classical limit of the open 
string spectrum on the D2-brane. This quantity can be obtained by 
analyzing the corresponding following \emph{open-string Laplacian}, 
\bea
\Delta\o \ = \  
- \frac{1}{e ^{-2\Phi_{os} }\sqrt{\det
  G_{os} }}\partial _\mu e ^{-2\Phi_{os} }\sqrt{\det G_{os} }\ 
  G^{\mu \nu}_{os} \partial_\nu\ \ .
\label{osl}
\eea
Note that $\Delta_{os}$ differs from the closed string Laplacian $\Delta$ 
in that it involves the open string metric $G_{os}$ and dilaton $\Phi_{os}$. 
We can determine the latter from the explicit formulas for the metric 
$g$ and the dilaton $\Phi$ of the cigar geometry. For the open string 
metric $G_{os}= g - F {g}^{-1} F$ we find  
\bea 
ds^2_{os} \ = \ \frac{k}{2}\frac{1}{\cosh^2\rho-\sin^2\s}\, 
  \left(\cosh^2\rho\, d\rho^2+\sinh ^2\rho\, d\theta ^2\right)\ \ . 
\eea
Similarly, the open string dilaton is given by 
\bea
e^{\Phi_{os}} \ = \ e ^\Phi\, 
 \sqrt{\det (g +F) g^{-1}}\ \propto\ \frac{1}{\sqrt{\cosh^2\rho -
 \sin^2\s}}\ \ .
\label{osdd2} 
\eea
Inserting both expressions into eq.\ (\ref{osl}), the open string 
Laplacian takes the form 
\bea
- \frac{k}{2}\Delta_{os} \ =\ \frac{2}{\sinh
  2\rho}\, \p_\rho\tanh\rho\ (\cosh^2\rho-\sin^2\s)\p_\rho
+\frac{\cosh^2\rho-\sin^2\s}{\sinh^2\rho}\, \p_\theta ^2\ \ .
\label{osld2}
\eea
The spectrum of this operator is the semi-classical limit of the
open string spectrum on a D2-brane. As usual, the corresponding 
spectral density can be constructed from the reflection amplitude, 
i.e.\ from the behavior of eigen-functions under the reflection 
$ j \rightarrow - j-1$. Let us only quote the results of this 
investigation: it turns out that the spectrum contains all values
$\Delta^j = -j(j+1)/k+ n^2/4k$ with $j=\j$, and that the 
$\s$-dependence of the reflection amplitude is given by
\bea
R(j,n|\s,\s)\pp \ \propto\  (\cos ^2\s)^{2j+1}\ \ .
\label{refd2limit}
\eea 
This information suffices to find the relative spectral density 
of the system. We also note that, for obvious geometric reasons, we 
do not expect to find open string states with non-zero winding number 
in the spectrum.

\subsection{The exact solution}

Concerning the 2-dimensional branes we claim that there exists a
1-parameter family of exact solution. Just like the branes in our
semi-classical discussion, the exact solutions are parametrized 
by a real parameter $\s$ which now can take values in the 
interval 
\bea
 \s \ \in \ [\, 0\, ,\, \frac{\pi}{2}(1+b^2)\, [\ \ . 
\label{d2inter} 
\eea
Note that  this interval shrinks to its semi-classical analogue
as we send $b$ to zero. For the associated one-point functions 
of closed string modes in the presence of a D2-brane we propose   
\bea
\langle \Phi^j_{nw}(z,\bar z) \rangle_\s^{\rm D2} & = & 
 \, \delta_{n,0}\, \N'(b) 
 \left(\frac{\Gamma(-j+\frac{kw}{2})}{\Gamma(j+1+\frac{kw}{2})}
 \ e^{i\s(2j+1)}+\frac{\Gamma(-j-\frac{kw}{2})}
  {\Gamma(j+1-\frac{kw}{2})}\ e^{-i\s(2j+1)}\right)\, \times
\nonumber \\[3mm]
& & \hspace*{.5cm} \times \ (k/2)^{\frac{1}{4}}
\, \Gamma(2j+1)\, \Gamma(1+b^2(2j+1))\, 
 \nu_b^{j+\half}\ \frac{1}{|z-\bar z|^{h^j_{0w}+\bar h^j_{0w}}}
\label{1ptd2}
\eea
where $\N'(b) = (8b^2)^{-1/4}$ is the same factor as for D1-branes. 
This formula holds for closed string modes from the continuous series.
It also encodes all information about the couplings of discrete modes, 
but they have to be read off carefully because of the infinite factors
(see the discussion in the case of D0-branes).
\medskip 

The construction of the open string partition functions or rather 
of their spectral densities uses the same objects that we introduced 
in our discussion of the D1-branes. We are able to write down  a 
consistent spectrum of open strings stretching between two D2-branes 
of parameters $\s,\s'$ only if they satisfy the condition
\bea
\s-\s'\ =\ 2\pi\, \frac{m}{k-2}\ \ , \ \ m\in \Z\ \ .
\label{qd2quant}
\eea
This is a $k$-deformed version of the semi-classical condition 
(\ref{qd2clas}). We claim that under the above condition, the open 
string spectrum between the branes $\s$ and $\s'$ contains the same 
discrete open string states as open string spectrum of one D0-brane 
of parameter $m$ (see eq.\ \ref{specd0}), and continuous states with 
relative spectral density
\bea
\Delta \SN(P,n|\s,\s')\ =\ \Delta \SN\left(P\left|i\frac{\s+\s'}{2},
  i\frac{\s+\s'}{2}\right.\right)+ \Delta \SN\left(P\left|i
 \frac{\s-\s'}{2}, i\frac{\s'-\s}{2}\right.\right)\ \ .
\label{specd2}
\eea
The relative density $\Delta \SN(P,n|\s,\s')= \Delta \SN(P,0|\s,\s')$ is 
the same as for open strings with $w \in \mathbb{Z}$ on a D1-brane
but with the real brane parameters replaced by purely imaginary 
ones. In formulas we claim that\footnote{Before version 3 of this
paper, there was a sign mistake in this formula: the contribution
$Z^{\Dz}_{mm}(q)$ appeared with a $+$ sign. This was due to a sign
mistake in the coefficient $a_{III}$ (\ref{aiii}) which is now
corrected. We should however not hurry to conclude that the
corresponding D2-branes configurations
are inconsistent. Indeed, we computed relative partition functions. We
can therefore make no claims about absolute signs of state multiplicities.
We would like to thank Dan Israel, Ari Pakman and Jan Troost for
bringing this mistake to our attention.
}
\bea \label{pfd2} 
 Z^{\D2}_{\s\s'}(q) \ = \ \int dP\ \sum_{n \in \Z} \ \Delta 
    \SN(P,n|\s,\s') \ 
    \chi^c_{(\j,n)}(q) \ - \ Z^{\Dz}_{mm}(q)
\eea
where $m = (\s-\s')/2\pi b^2 \in \Z$. Note that the density of continuous
open string states diverges when $\frac{\s+\s'}{2}$ reaches the upper 
bound $\halfpi (1+b^2)$ of eq.\ (\ref{d2inter}). Based on the explicit 
formulas for the open string spectral density we are now able to  
identify the parameter $\sigma$ that enters the exact solution with the 
parameter $\s$ in our semi-classical analysis. To this end we evaluate 
the exact formula for the spectral density in the semi-classical limit 
$ b \rightarrow 0$ and compare the result with the semi-classical 
density obtained from eq.\ (\ref{refd2limit}). The details are left 
to the reader.

\subsection{D2-branes from descent }

We obtain the D2-branes by rotating the Euclidean \AA-branes in 
$\Hp$ and then descending to the cigar (see figure \ref{76cigd2ads}). 
The relevant rotation matrix $U$ can be found \cite{pst} and the 
corresponding one-point function are easily worked out, 
\bea \langle \Phi^{{\rm H},j}_{np} (z,\bar z) \rangle_{r} \ = \ 
 \N'(b)\,  \delta_{n,0}\  \left[C'_j(p)\, e^{-r(2j+1)}+
  C'_j(-p)\, e^{r(2j+1)}\right]\ 
  \frac{\nu_b^{j+\half}\Gamma(1+b^2(2j+1))}
  {|z-\bar z|^{2\Delta_j}} \label{1ptD2}\ \ , 
\eea
where the function $C'_j(p)$ is defined through
$$ C'_j(p) \ = \ \int_0^1 dy\ y^{-2j-1+ip} |y^2-1|^{2j} =
\frac{\Gamma(2j+1)\Gamma(-j+ip/2)}{2\ \Gamma(j+1+ip/2)} . $$ 
Once more we multiply this by the one-point functions (\ref{1ptN}) of 
a free boson with Neumann boundary condition and insert $p = - i k w$ 
to obtain our formula (\ref{1ptd2}) for the bulk one-point function of 
a D2-brane. The \AA brane parameter $r$ is related to the parameter 
$\s$ in the 2-form field strength (\ref{fd2}) on D2-branes via $r=i\s$. 
Let us mention that we could  introduce a second parameter by shifting 
the branes in \H\ along the $\tau$ direction. This extra freedom is 
associated with turning on a Wilson line. Since it enters simply as  
$w$-dependent phase, we will not consider this any further.  
\smallskip 

The D2-branes on the cigar can be considered as non-compact analogues
of the B-branes in the coset \SUU\ (see \cite{MMS}). In fact, 
the $U$-rotated Euclidean \AA branes in \H\  from which we descended 
to our D2-branes may be Wick rotated into \Ht branes in \AAA. The 
latter are localized along the surfaces 
\bea \label{H2rel}  
\cosh \rho \ \sin t \ = \ \sin \sigma\ \ ,   
\eea 
i.e.\ along conjugacy classes of \SL. Notice that D2-branes with $\cosh \rho 
\geq \beta$ which cover only part of the cigar should likewise be related to 
the unphysical $dS_2$ branes in \AAA\ \cite{bp}. 
\smallskip 

\Fig{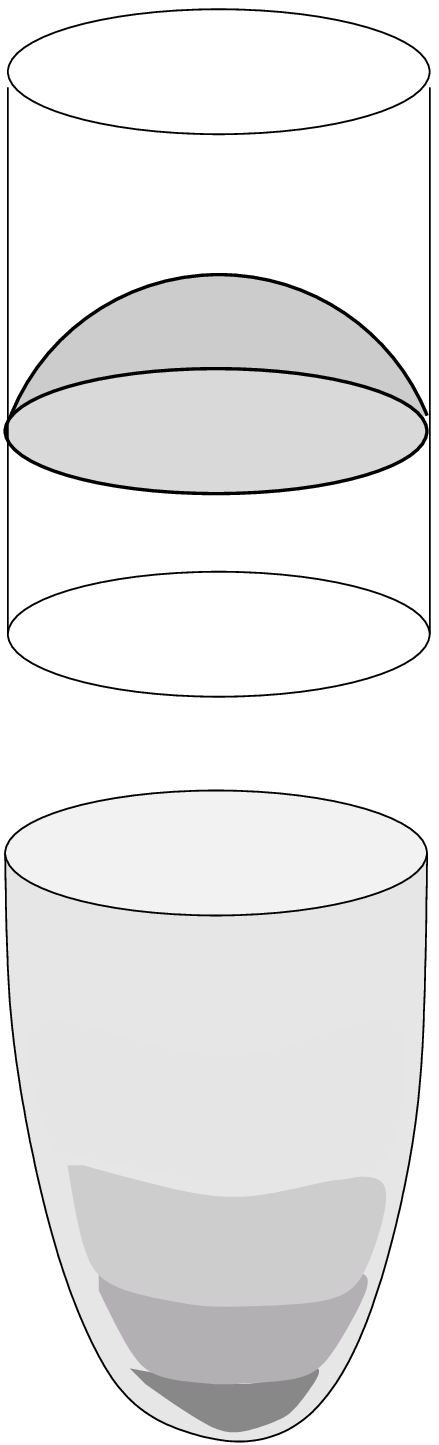}{3}{A D2-brane in the cigar descending from an
  \Ht brane in \SL\ or \H }

\subsection{Cardy consistency condition}

Our last task here is to evaluate the annulus amplitude between 
two D2-branes of parameters $\s$ and $\s'$ from the boundary 
states (\ref{1ptd2}) and to compare the answer with our 
expression (\ref{specd2}) for the open string spectrum. 
Using the couplings $\Psi_\s$ 
in the boundary states, we can compute
\bea
Z^\D2_{\s\s'}(\tilde q)\ =\ \int dj\ \sum_{w}\ \frac{
\Psi_\s(j,w) \Psi_{\s'}(j,w)^*
}{\la \Phi^j_{0w} 
  \Phi^j_{0w} \ra }\ \chi_{(j,\frac{kw}{2})}(\tq)\ \ .
\label{z1pt}
\eea
As it stands, this is just a formal expression in which the 
`integral' runs over both continuous and discrete series. The 
two-point function in the denominator is trivial in the case 
of the continuous series but is has a non-trivial dependence 
on the labels of discrete closed string states. We can spell 
out the integrand of eq.\ (\ref{z1pt}) explicitly, in a form 
that is valid for discrete and continuous values of the spin 
$j$, 
\begin{multline}
\frac{
\Psi_\s(j,w) \Psi_{\s'}(j,w)^*
}{\la 
  \Phi^j_{0w} \Phi^j_{0w} \ra }\ =\ 
\frac{\pi ^2 \sqrt{kb^2}}{\sin 2\pi j\ \sin \pi b^2 (2j+1) } \times
\\[2mm] 
\times
\left[
2\cos (2j+1)(\s+\s')+\frac{\sin\pi(j+\frac{kw}{2})}{\sin
  \pi(j-\frac{kw}{2})} e^{-i(2j+1)(\s-\s')} + 
\frac{\sin\pi(j-\frac{kw}{2})}{\sin
  \pi(j+\frac{kw}{2})} e^{i(2j+1)(\s-\s')}
\right]\ \ .
\label{d2disc}
\end{multline}
In order to modular transform the annulus amplitude (\ref{z1pt})
we employ simple trigonometric identities and rewrite the previous 
formula for the coefficients as
\bea
\frac{
\Psi_\s(j,w) \Psi_{\s'}(j,w)^*
}{\la \Phi^j_{0w}
  \Phi^j_{0w} \ra } \ = \ 
\frac{2 \pi ^2 \sqrt{kb^2}}{\sin \pi b^2 (2j+1) }\ \left\{ a_{I}+ a_{II}+
 a_{III}+a_{IV} \right\} \ , 
\label{fourterms}
\eea
where the four terms in the brackets are given by
\bea
a_{I} &=& \frac{\cos(2j+1)(\s+\s')}{\sin 2\pi j}  \\[2mm]
a_{II} &=& \cos(2j+1)(\s-\s')\ \coth 2\pi j   \\[2mm]
\label{aiii}
a_{III} &=& -\cos(2j+1)(\s-\s')\ \frac{\sin 2\pi j}{\cos 2\pi j - \cos
  \pi k w }  \\[1mm]
a_{IV} &=& i\ \sin (2j+1)(\s-\s')\ \frac{\sin \pi kw}{\cos 2\pi j - \cos
  \pi k w } \ \ .  
\eea
Let us note that the fourth term $a_{IV}$ does not contribute to the 
annulus amplitude because it is odd in $w$. The three other terms, 
however, are real and non-vanishing. Contributions from the discrete
closed string states are encoded entirely in the third term $a_{III}$.  
In fact, discrete states can only contribute through singularities of 
the integrand (see subsection \ref{cardyd0}) and the terms $a_I$ and 
$a_{II}$ are finite for physical values of $j$ in the discrete 
series\footnote{This may not always be true when $k$ is rational, in which
  case half-integer values of $j$ may appear in the spectrum. However,
  the positions of the corresponding poles of $a_{I}$ and $a_{II}$
 are then independent of $w$, and we believe that they should not be
 taken into account.}. 
The term $a_{III}$, on the other hand, is infinite when $(j,w)$ fall 
into the set of physical discrete states. The associated contribution 
to the annulus amplitude is given by the residue of the pole. With 
this in mind it is not hard to see that the term $a_{III}$ furnishes 
both the continuous and the discrete closed string modes' contributions 
to the annulus amplitude of a D0-brane with parameter $m=\frac{k-2}{2\pi}
(\s-\s')$, provided this $m$ is an integer (see equations (\ref{specd0},
\ref{cyld0disc})). These contributions come with a minus sign.
We thus recover the announced discrete part of 
open string spectrum on D2-branes along with a quantization condition
for the difference $\s - \s'$ of the branes' parameter. When the latter 
is  not obeyed, we cannot interpret our amplitude as an open string
partition function. 
\smallskip 

We still have to deal with terms $a_I$ and $a_{II}$. Their contribution 
comes entirely from continuous closed string states and it is rather 
easy to modular transform. Indeed, the integral over $P$ and the sum 
over $w$ decouple and we find continuous open string modes with 
density (\ref{specd2}). Note that the computation gives a well-defined 
density only if the mean $(\s+\s')/2$ of the branes' parameters $ \s$ 
and $\s'$ belongs to the interval (\ref{d2inter}). In particular, the 
parameter of any single brane has to belong to the same interval so 
that we obtain an exact version of the classical bound on $\s$. This
concludes the Cardy check for D2-branes in the cigar.

\setcounter{equation}{0} 
\section{Conclusion and open issues }

In this work we have presented a complete analysis of (maximally
symmetric) branes in the 2D black hole background. The main results
include exact formulas (\ref{1ptd0}), (\ref{1ptd1}), (\ref{1ptd2}) 
for the various boundary states and the boundary partition functions 
(\ref{specd0}), (\ref{pfd1}), (\ref{pfd2}). To write down the latter 
we have also presented formulas for the boundary reflection amplitudes, 
i.e.\ for the 2-point functions of boundary fields. It would certainly 
be worthwhile working out formulas for the spectrum of open strings 
that stretch in between branes of different dimension. At the moment 
we only have expressions for open strings stretching between D0 and 
D2-branes. 
\smallskip 

The most important information, however, that is missing for a 
complete solution of the boundary theories are the various boundary 
3-point functions. Unfortunately, these are not known even for the 
\H theory and so finding explicit formulas for the 3-point couplings 
remains an interesting open problem for future research. Its solution 
would be a crucial step toward studying the dynamics of extended branes
on the cigar (see \cite{TeschLD} for related work on Liouville theory).  

Here we can only formulate a conjecture on brane dynamics. The 
first one concerns systems of D0-branes and a D2-brane (see also 
\cite{bfpr}). Given the identification of the D2-brane parameter 
$\sigma$ with a D0-brane charge (see last section) it is tempting 
to conjecture that the bound state between $N$ D0-branes with $m=1$ 
and a single D2-brane of parameter $\s$ results in a single D2-brane 
with parameter $\s' = \s + 2\pi b^2 N$. In other words, we imagine 
a process in which a D2-brane absorbs a single D0-brane or a finite 
number thereof, as in Figure \ref{75cigd2}. Note that our identification 
of the bound state  assumes that the parameters stay in the `physical' 
range $\s +2\pi b^2 N < \pi(1+b^2)/2$ and we do not have any candidate 
boundary state for the condensate if this bound is exceeded.  
It would 
certainly be interesting to verify these this proposal through  
a perturbative analysis in the exact boundary CFT (see \cite{FrSc,
Fre} for similar studies in the case of compact parafermions).
\medskip 
 
There exist several other issues that deserve further investigation. 
The most interesting involve the D0-branes at the tip of the cigar. 
We have pointed out at several places throughout the paper that the D0 
and D2-branes are non-compact analogues of the B-branes in \cite{MMS}. 
In the compact case, B-branes have been used to construct symmetry 
breaking branes on the group manifold \SU. These are either 1- 
or 3-dimensional. A similar construction (see \cite{QueSch} 
for a more systematic treatment) can be applied to the D0-branes 
in the 2D black hole and it provides us with a brane that stretches 
out along the 1-dimensional line $\rho =0$ in \H. 
%SR : minor changes in the following
The existence of 
such symmetry-breaking branes in \SL\
can also be inferred from a semi-classical analysis \cite{gor,thomas}, 
but their exact construction was not known before (see \cite{Raj} 
for some attempts in this direction). Volume filling branes in \H can 
be obtained similarly from the D2-branes in the black hole.  
\smallskip 

Another direction involving primarily the D0-branes concerns 
the generalization of recent developments in 2-dimensional 
string theory and the $c=1$ matrix model. It was argued in 
\cite{McGVer1,Mart,KlMaSe} that the well known matrix dual 
of Liouville theory may be interpreted as the effective 
field theory describing the dynamics of localized branes in 
Liouville field theory. The dual matrix model for the 2D 
black hole background has been found in \cite{KaKoKu} and, 
in complete analogy to the Liouville case, it should describe 
the dynamics of our D0-branes \cite{akk}. Higher dimensional generalizations
of this duality, which all contain the 2D black hole as a building 
block, arise in the context of little string theory. The results 
we have described above allow to construct various branes in the 
dual 9+1-dimensional string backgrounds. 
\medskip     

%SR: removed last paragraph. 

\Fig{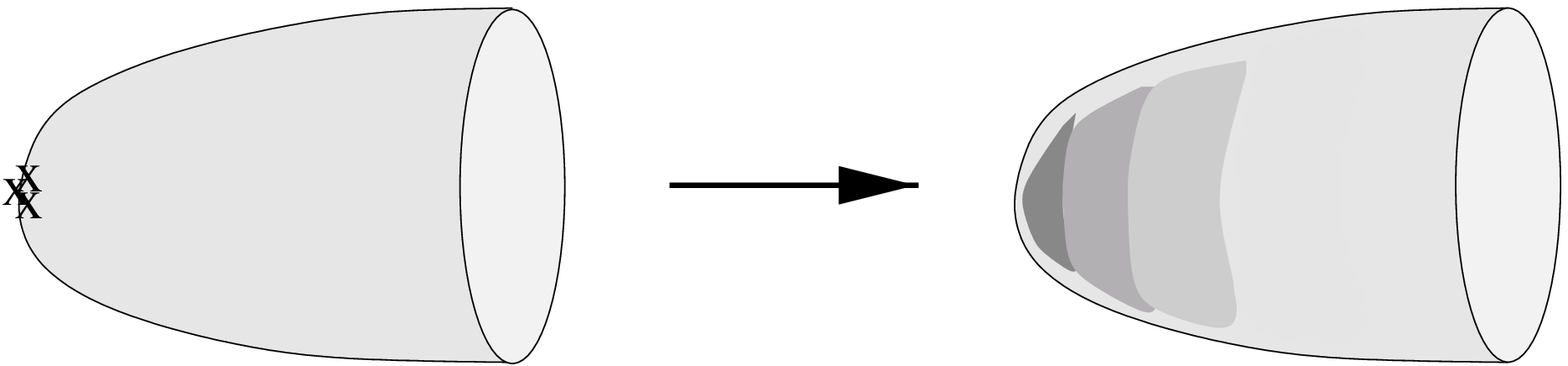}{12}{D0-branes being absorbed by a D2-brane }
\vspace{3truemm}

\noindent 
{\bf Acknowledgements:}
J.\ Teschner and S.\ Fredenhagen contributed in the initial phase 
of this project and it is a pleasure to thank them for many 
helpful discussions. We are also grateful to C.\ Bachas, A.\
Fotopoulos, A.\ Giveon, 
M.\ Petropoulos, T.\ Quella, A.\ Recknagel, and A.\ Schwimmer for 
their interest and useful remarks. This work has been started while 
one of the authors (S.R.) was visiting the Albert-Einstein Institut 
in Golm and it was mainly carried out at the CPHT of the Ecole 
polytechnique. We thank both institutions for their support and a 
stimulating atmosphere. Support for this work also came from the 
EC network grant ``EUCLID'', contract number HPRN-CT-2002-00325.

\begin{appendix} 
\section{Characters of discrete representations } 
\setcounter{equation}{0} 

In this appendix we derive the formula (\ref{cardisc}) that we 
use for the characters of the discrete series on the cigar. Our 
strategy is to insert the expression (\ref{Hcardisc}) for 
characters of discrete \SL\ representations into relation 
(\ref{dcharint}) and to evaluate the integral explicitly. Before we 
go into this computation, let us make a few comments on the 
function $\vt(q,z)$. It is defined by    
\ber
\vt(q=e ^{2\pi i\tau},z=e ^{2\pi i\theta}) &=& -2q^{1/8} \sin \pi
\theta\,  \prod_{n=1}^\infty \, (1-q ^n)(1-zq^n)(1-z^{-1}q^n) \\
&=& \sum_{m=-\infty}^\infty (-1)^{m}q^{\half(m-\half)^2} z^{m-\half}
\eer
and enjoys the following behavior under modular transformation
$$ \vt(q,z) \ = \  -(-i\tau)^{-\half}\, e^{-i\pi \theta ^2/\tau}
\, \vt(\tq,\tz) $$
where $\tq = \exp(-2\pi i/\tau)$ and $\tth = \theta/\tau$. 
\smallskip 

The computation of the integral (\ref{dcharint}), and therefore 
the proof of formula (\ref{cardisc}), is easily seen to boil 
down to the calculation of the following contour integrals
for $\kappa \in \Z$,   
\bea
\frac{1}{2\pi i} \int \frac{dz}{z} \frac{z^\kappa}{1-z}\, 
  \frac{1}{\Pi_{n=1}^\infty (1-zq^n)(1-z^{-1}q^n)}
\eea 
over the circle $|z| = 1 -\epsilon$ (recall that we displaced 
the contour in eq.\ (\ref{dcharint}) slightly to regularize the 
integral). If $\kappa \geq 1$ then all singularities of the 
integrand inside the circle come from the first order poles 
at $z = q^m$. Hence, the contour integral can be evaluated 
using that in the vicinity of $z = q^m$ we have 
\bea
\frac{1}{1-z} \frac{1}{\Pi_{n=1}^\infty (1-zq^n)(1-z^{-1}q^n)}\  
\simeq \ 
\frac{q^m}{z-q^m}\, \frac{q^{\frac{1}{12}}}{\eta(q)^2} 
(-1)^{m+1} q^{m(m-1)/2}
\ \ . 
\eea
If $\kappa \leq 0$, on the other hand, we move the integration 
contour to infinity, picking up contributions from the poles at 
$z = 1 $ and $z = q^{-m}$, all decorated with an extra minus sign 
to take the orientation of the contour into account. The results 
of these manipulations agree with our expressions (\ref{cardisc}) 
for the discrete coset characters. 

\section{Cardy computation for \S-branes in \H }
\setcounter{equation}{0} 

Our aim here is to extend the proof of world-sheet duality 
for the \S-branes in \H (see \cite{pst}). In contrast to the 
calculations performed in \cite{pst}, we will not only check the 
world-sheet duality for the usual annulus amplitude, but for the
more interesting quantity,
\bea
Z_{mm'}^{\S}(q,z)\ =\ \Tr q^{L_0-\frac{c}{24}}\ z^{J^0_0}\ \
\eea
The operator  $J^0_0$ denotes
the zero mode (in the sense of affine Lie algebras) of the \SLC\ 
current $J^0$ of the \H\ sigma model. In our analysis we compute 
the quantity $Z_{mm'}^{\S}(q,z)$ from the one-point functions 
(\ref{1ptd0h3}) of the \S branes in \H and express the result 
in terms of unspecialized characters,  
\bea
\tq^{\frac{k}{4}\theta^2}\, Z^{S^2}_{mm'}(q,z) & = & 
  \,^{\S}\!\!\!\langle m|\, \tq
^{(L_0+\bar{L}_0)-\frac{c}{24}}\,  e ^{2\pi i \tth J^0_0}\, |m'\rangle^{\S}
\nn \\[2mm]
&=& 
\int dPdp\ \Psi^{\S}_m\!\!(\j)\, \Psi^{\S}_{m'}\!\!(\j)^\ast \,  
\tq^{b^2P^2}\, e^{-\pi \tth p}\, \frac{(-2)\sin \pi \tth}{\vt(\tq,\tz)}
\nn \\[2mm] 
& = & 
- 2 \sqrt{2b^2} \int dP\ \frac{\tq ^{b^2P^2}}{\vt (\tq,\tz)}\, 
  \sum_J\,  \sinh 2\pi b^2 P (2J+1)\ \sinh 2\pi P \tth 
\label{intP}
\eea
where have inserted the eigen-values $i \frac{p}{2}$ of the operator 
$J^0_0$ and $\Psi^{\S}_m\!\!(j) = \langle \Phi^{{\rm H},j}_{np} 
(z=\frac{i}{2})\rangle_m$. In the last line the summation over 
$J$ runs over the same set as in eq.\ (\ref{specd0}). Now we 
are prepared to perform the modular transformation,   
\bea
Z_{mm'}^{\S}(q,z) & = & 
-2 \sum_J\, %\tq^{\frac{k}{4}\theta ^2}\, 
q^{-b^2(J+\half)^2} \, \frac{\sin \pi \theta (2J+1)}{\vt (q,z)} 
\label{zss}
\\[2mm]
&= & 
\sum_J \, %\tq^{\frac{k}{4}\theta ^2}\, 
 \left( \chi^{{\rm H},d}_J(q,z) - \chi^{{\rm H},d}_{-J-1}(q,z)
 \right)\ \  .
\label{specd0h3}
\eea
We thus find the same discrete spectrum as in \cite{pst}, now written
in terms of unspecialized characters (\ref{Hcardisc}). 
\end{appendix} 

%\bibliographystyle{JHEP-2}
%\bibliography{981cigbib}

\providecommand{\href}[2]{#2}\begingroup\raggedright\endgroup

\end{document}